\documentclass[aps,amsmath,amssymb,titlepage, twocolumn, nofootinbib]{revtex4-1}

\usepackage{graphicx}
\usepackage{nicefrac}
\usepackage{amsfonts}

\usepackage{amssymb}
\usepackage{amsmath} 
\usepackage{hhline}
\usepackage{subfig}
\usepackage{multirow} 
\usepackage{tabularx} 
\usepackage{array}

\usepackage{units}
\usepackage{color} 
\usepackage{tensor} 
\usepackage{braket}
\usepackage{comment}

\usepackage[mathscr]{euscript}

\usepackage{bm}
\usepackage{hyperref}

\newcommand{\grafe}[1]{\left\{ #1 \right\}}
\newcommand{\tonde}[1]{\left( #1 \right)}
\newcommand{\quadre}[1]{\left[ #1 \right]}

\makeatletter
\renewcommand\@makecaption[2]{%
  \par
  \vskip\abovecaptionskip
  \begingroup
   \small\rmfamily
    \begingroup
     \samepage
     \flushing
     \let\footnote\@footnotemark@gobble
     \@make@capt@title{#1}{#2}\par
    \endgroup
  \endgroup
  \vskip\belowcaptionskip
}
\makeatother

\begin{document}

\title{Fluctuation driven transitions in localized insulators: \\Intermittent metallicity and path chaos precede delocalization}

\author{Valentina Ros}
\email{valentina.ros@universite-paris-saclay.fr}
\affiliation{Universit\'e Paris-Saclay, CNRS, LPTMS, 91405, Orsay, France}

\author{Markus M{\"u}ller}
\email{Markus.Mueller@psi.ch}
\affiliation{Paul Scherrer Institut, CH-5232 Villigen PSI, Switzerland}
\affiliation{The Abdus Salam International Center for Theoretical Physics, Strada Costiera 11, 34151 Trieste, Italy}

\date{\today}

\begin{abstract}
We study how interacting localized degrees of freedom are affected by 
slow thermal fluctuations that change the effective local disorder. We compute the time-averaged (annealed) conductance in the insulating regime and find three distinct insulating phases, separated by two transitions. The first occurs between
a {\it non-resonating insulator} and an {\it intermittent metal}. The average conductance is always dominated by rare temporal fluctuations. However, in the intermittent metal, they are so strong that the system becomes metallic for an exponentially small fraction of the time.  A second transition occurs within that phase. At stronger disorder, there is a single optimal path providing the dominant contribution to the conductance at all times, but closer to delocalization, a transition to a phase with fluctuating paths occurs. This last phase displays the quantum analogon of configurational chaos in glassy systems in that thermal fluctuations induce significant changes of the dominant decay channels. While in the insulator the annealed conductance is strictly bigger than the conductance with typical, frozen disorder, we show that the threshold to delocalization is insensitive to whether or not thermal fluctuations are admitted.
This rules out a potential bistability, at fixed disorder, of a localized phase with suppressed internal fluctuations and a delocalized, internally fluctuating phase.
\end{abstract}
\maketitle

\section{Introduction}
Strong Anderson localization and its interacting analogue, Many-Body Localization (MBL), both arise due to the disorder-induced suppression of resonant couplings between close-by states in configuration space. Strong disorder renders typical energy differences between such configurations too large to be overcome by the off-diagonal tunneling terms that connect them. The sparse remaining resonances turn out to be harmless (at least in one dimension, in the case of many-body systems) so that localization, non-ergodicity and a resistance that grows exponentially with system size are preserved. For non-interacting quantum particles this has been understood already by Anderson~\cite{anderson1958absence}, and it has been argued at various levels of rigor for the many-body case over the last 15 years~\cite{Basko:2006hh,gornyi2005interacting,imbrie2016}, taking up and developing much further an initial perturbative analysis by Fleishman and Anderson~\cite{Fleishman1980interactions}, cf. also \cite{AbaninRecentProgress, parameswaran2017eigenstate,imbrie2017local, abanin2018ergodicity} for recent reviews. MBL has been argued to be the most robust route to achieve ergodicity breaking (without invoking spontaneous symmetry breaking), to suppress long range transport and to avoid thermalization in isolated interacting quantum systems \cite{huse2015review}.  

The above considerations hold for {\em typical} realizations of {\em static} disorder. It is well-known, however, that rare correlated realizations of disorder can nevertheless exhibit delocalization and finite transport, even though they are encountered with vanishing probability in the thermodynamic limit. Rare optimal configurations are also known to play an important role in the transport of insulators, as they dominate the elastic transmission through broad junctions~\cite{LifshitzKirpichenkov} and they affect inelastic hopping processes, too. \cite{pollak1973note, levin1988transverse,tatarkovskii87}

Here we go beyond this type of analysis by asking about the role of {\em temporal} fluctuations. The dominance by rare temporal fluctuations that we will find is nevertheless reminiscent of the dominance of rare (static) transmission channels in broad junctions, whereby different temporal realizations of the effective disorder now take the role of the spatial position of conduction channels,  the temporal average replacing the spatial average. In a higher-dimensional situation, temporal and spatial inhomogeneity will be important simultaneously. As we will see, their interplay gives rise to a new regime where the spatially dominant channel fluctuates itself in time.  

We point out that temporally fluctuating disorder can induce delocalization, even if at every instance of time the disorder realization would be classified as localizing, if it were static. This  is due to the creation of semi-local resonances at different moments in time.  This phenomenon has  been studied, e.g., in the context of periodically modulated disordered Hamiltonians~\cite{gopalakrishnan2015low, abanin2016theory}: The slower the variations the higher is the danger to encounter resonances over long enough time windows during which adiabatic state changes take place, which delocalize the system in the long run and restore ergodicity. This mechanism poses a fundamental problem when one tries to utilize disorder-induced many-body localization to protect topological order and anyon braiding at finite temperature.~\cite{Sondhi2015} Ovadyahu has used this observation in Refs.~\cite{Ovadyahu2021} to study the reach of long range resonances as a function of the excitation state of a disordered electron glass. Those experiments suggest that the fluctuations in a sample at equilibrium are very slow and infrequent (in contrast to an excited sample), an assumption that we will make for our theoretical analysis.

In systems with a conserved charge, a time dependence of the disorder almost invariably increases the time-averaged conductance associated to the charge. Indeed,  since the conductance is an exponentially small quantity, its time average is likely to be dominated by rare temporal fluctuations during which the conductance increases exponentially, even if only for an exponentially small fraction of the total time. It is often the first exponential  that dominates, as we will show with an explicit calculation in this work. Likewise, the life time of excitations localized in the bulk of a sample is likely to be limited by such rare fluctuations, rather than by the decay in the presence of a typical disorder configuration. 

This then raises an interesting question: 
Anderson's criterion for the  breakdown of localization  requires that  the decay rate to infinity turns from being exponentially suppressed in the system size to becoming finite. Since it turns out that in the presence of fluctuations these decay rates depend on whether one takes their annealed, i.e. time-averaged  value, or their quenched value in a static, typical configuration one might surmise that the delocalization transition of the quantum system actually depends on whether or not the disorder is fluctuating {(which strictly speaking assumes a bath, as we will discuss in more detail below)}. 
From this it is then a natural next step to ask what would happen if fluctuating effective disorder arose not from an external source, but were generated internally, by the dynamics of the system itself {(which is consistent only within a many-body-delocalized phase)}. If the delocalization transition  indeed hinged on the presence or absence of fluctuations, the instability would actually depend on whether one approaches the transition from the delocalized or the localized side. This would then suggest a region of bistability where either phase would be self-consistent - a scenario that would be in stark contrast with the currently favored scenario that the many-body delocalization transition occurs at a unique, well-defined critical point.~\cite{thiery2018many, dumitrescu2019kosterlitz,morningstar2020many}

The question about the effect of fluctuations is particularly relevant in systems where interactions have the predominant role of tuning the effective disorder.
This differs from their role in the canonical models of weakly interacting disordered quantum particles that were studied in the wake of MBL~\cite{Basko:2006hh,gornyi2005interacting}. In those cases the prime role of interactions is to tune the number of scattering channels that allow for long range transport. 
Here instead we are interested in systems of particles or spins, where the interaction terms act mostly ``classically", in the sense that they commute with each other and with the disordered potential part of the Hamiltonian; a simple example are density-density interactions of strongly localized electrons or Ising interactions of spins in random longitudinal fields. In such models, transport is mainly due to kinetic hopping or spin-flip terms which compete with the interaction-induced potential landscape. The interaction terms thus strongly affect the effective local energy spectrum that excitations encounter as they propagate. Note that the net effect of interactions in this framework is not obvious from the outset:
On the one hand, thermal fluctuations of the degrees of freedom with which an excitation interacts (other spins or electrons) may generate effective local fields that are more resonant with the considered excitation, enhancing small denominators and thus favoring delocalization. 
In certain cases, on the other hand, the interactions may even 
enhance the localization tendency with increasing temperature, because thermal configurational disorder translates into an increased width of the disorder distribution~\cite{KaganMaksimov, SchiulazVarmaMueller,ros2017remanent, Huveneers}. 
``Classical" interactions (i.e. interactions diagonal in the natural, localized basis), may also play a significant role for the usual channel of many-body delocalization, in particular if they are longer ranged. In that case a  flipping degree of freedom can bring two different degrees of freedom into resonance with each other and thereby kick off a side avalanche of decay. This phenomenon of spectral diffusion~\cite{BurinKagan95, gornyi2017spectral} tends to decrease the stability of the MBL phase. Here, we restrict ourselves to short ranged models where such side-avalanches can be neglected.

The localization properties of one of  the simplest realizations of  a system of predominantly classically interacting degrees of freedom was studied in Ref.~\onlinecite{cuevas2012level} and analyzed on a Bethe lattice for simplicity, as a proxy for finite dimensional lattices with a site connectivity larger than $2$. Despite the fact that the distribution of effective local fields (sampled over all sites) remained temperature-independent in that model, the results of Ref.~\onlinecite{cuevas2012level} suggested that the presence of thermal fluctuations shifted the localization phase boundary towards stronger disorder. As we mentioned above, if true, such an effect would entail the possibility of an intermediate range of disorder strength where two physically different phases would be self-consistent and locally stable: (i) a non-fluctuating, more strongly localized non-thermal phase with frozen effective potentials, and (ii) an ergodic delocalized phase with thermally fluctuating effective potentials. 

Here we revisit this intriguing scenario. Our analysis shows, however, that such a bistability of localized and delocalized phases is impossible: while temporal fluctuations of local potentials definitely affect the spatial structure of localized excitations and their effective localization length, the localization phase boundaries (or the associated crossovers) will be shown to be independent of whether or not thermal fluctuations are included in the analysis. This is so despite the fact that analytically continuing the
annealed (``fluctuating") conductance  from the strong disorder regime  would actually predict delocalization at stronger disorder than in a disorder-quenched system. However, we will show that a phase transition inside the insulating regime (from a deep, ``non-resonating insulator" to an ``intermittent metal") invalidates the analytical continuation and thus the prediction of a shifted delocalization transition.
{ Non-resonating insulator and intermittent metal exhibit qualitatively different average conductances, with a potentially observable transition between them}~\cite{Krinner2017}. In the intermittent metal the conductance is dominated by rare thermal fluctuations that induce metallic behavior, while the sample still looks well insulating at typical instants. 
 
We will find that in dimensions $d>1$ thermal fluctuations induce a further transition within the intermittent metallic phase. That second transition is a quantum analogon of the freezing-unfreezing transition occurring in certain models of glasses \cite{derrida1988polymers,Derrida:1981jd}. While in static (or relatively weakly fluctuating) disorder the conductance is dominated by the propagation along one (or very few) rather well-defined paths through the sample, this changes in the vicinity of delocalization, where the dominant path starts to fluctuate with the thermal fluctuations of the effective local disorder. This is closely related to the well-known ``chaos" phenomenon in glassy systems where the ground state configuration often changes in a chaotic manner as the disorder potential is modified.~\cite{McKayChaos,bray1987chaotic, CrisantiRizzoChaos}

{ We reach these results as follows.}  In Sec.~\ref{sec:Model} we introduce our model and recall how to compute approximately the decay rate of local excitations. In Sec.~\ref{sec:Averages} we discuss how to account for the  thermal fluctuations induced by a weak coupling of the system to a bath. We discuss the limits of validity of our approach and comment on the implications of our results for the localization-delocalization transition, if the fluctuations are interpreted as being generated internally by the system in a delocalized phase. The simpler one-dimensional case in which one single decay path is accessible to the system is analyzed in Sec.~\ref{sec:FrozAnn}, to illustrate how the transition between the non-resonating insulator and the intermittent metal occurs. These results are generalized to many decay paths in Sec.~\ref{sec:ManyPaths}, where we will find the path-unfreezing transition. In Sec.~\ref{sec:Discussion1} and \ref{sec:Discussion2} we discuss the physical interpretation of our results. The conclusions are given in Sec.~\ref{sec:Conclusions}.

\section{Model and locator approximation}\label{sec:Model}
As motivated above we consider models where the interactions predominantly shape the effective energy landscape. We focus on spin systems on a lattice, with Hamiltonian of the generic form: 
\begin{equation}\label{eq:HamCuevas}
H= H_{\rm cl}(\{\sigma_i^z\})
- \sum_{\langle i,j \rangle} J_\perp \tonde{ \sigma^+_i \sigma^-_j+ \sigma^-_i \sigma^+_j},
\end{equation}
where $H_{\rm cl}$ is a function of the  classical Ising spin variables $\sigma^z_j$ only, and thus only contains mutually commuting terms.
Quantum fluctuations and dynamics arise through the spin-flip term with amplitude $J_\perp$. The notation $\langle i,j \rangle$ indicates that the sites  $i,j$ are nearest-neighbors in the lattice.

\begin{figure}[ht]
    \includegraphics[scale=.5]{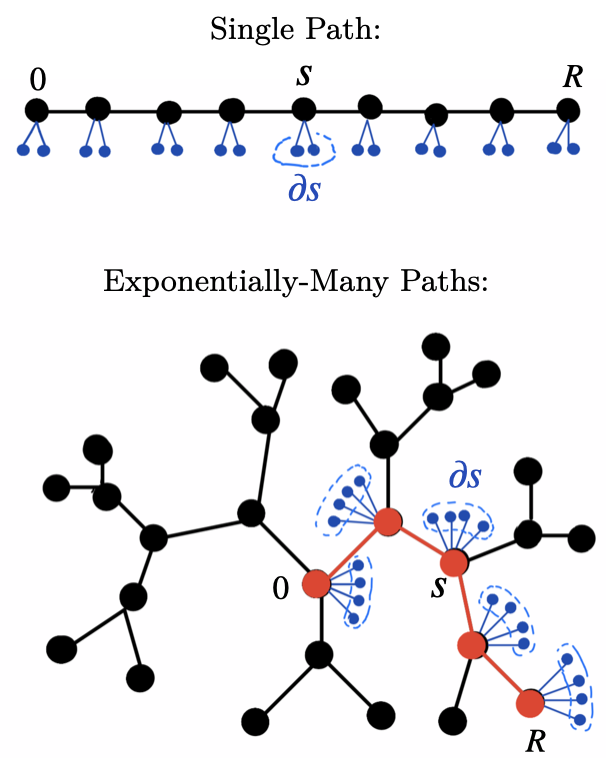}
    \caption{Schematic representation of the set-ups discussed in this work. The top figure shows sites $s$ belonging to a path of length $R$, each site being connected to other $N$ ``environmental"  spins (here $N=2$) belonging to the neighborhood $\partial s$. The bottom figure shows in red one out of exponentially many paths  of length $R$ on a tree with branching number $k=2$. {Each site $s$ of the tree is connected to $N \gg 1$ environmental spins in $\partial s$.}}\label{fig:Pictorial}
  \end{figure}

The effective local field seen by spin $i$ is  
\begin{equation}
 h_i^{\text{eff}}  = -\frac{\partial H_{\rm cl}}{\partial \sigma_i^z},
\end{equation}
which for pairwise Ising interactions takes the form
\begin{equation}
 h_i^{\text{eff}} =  \epsilon_i  + \sum_{j \in \partial i} J_{ij} \sigma^z_j \equiv  h_i^{\text{eff}} \tonde{\epsilon_i, \vec{\sigma}_{\partial i}},
\end{equation}
where $\vec{\sigma}_{\partial i}$ denotes the configurations $\sigma^z_j$ of the spins in the neighborhood $\partial i$ of  $\sigma^z_i$, see Fig. \ref{fig:Pictorial}.  We assume the random fields $\epsilon_i$ to be independent on every site, with an identical distribution $f(\epsilon)$ with width $W$. Models of the form \eqref{eq:HamCuevas} arise rather ubiquitously in the theory of disordered quantum magnets~\cite{Kucsko2018, gornyi2017spectral}, quantum Coulomb glasses~\cite{epperlein1997quantum, li1993effect, vignale1987quantum,amini2014multifractality}, disordered supersolids~\cite{SuperglassesBiroli2008,yu2012meanfield,carleo2009bose}, cold atomic systems~\cite{MMStrackSachdev}, or disordered superconductors \cite{sacepe2011localization, Feigelman:2010fy}.

The nearly classical variables $\sigma^z_i$ have a dynamics induced by the transverse term. {The local fields $h^{\rm eff}_i$ evolve with time due to the fluctuations of the neighboring spins, which we suppose to be induced by a very weak coupling to a bath, e.g. of phonons.}
If the neighboring spins fluctuate thermally, the probability to see a field $h_i$ at a given site $i$ is given by
{\medmuskip=0mu
\thinmuskip=0mu
\thickmuskip=0mu
\begin{equation}
\label{Ps}
P(h_i| \vec{\epsilon}_{ i}) = \hspace{-.3 cm} \sum_{\{\sigma_j^z\}|{j \in \partial i}} \hspace{.1 cm} \prod_{j \in \partial i}  \frac{e^{-\beta \epsilon_j \sigma_j^z}}{2 \cosh (\beta \epsilon_j)}\delta \tonde{h_i - \epsilon_i  - \sum_{j \in \partial i} J_{ij} \sigma^z_j}.
\end{equation}
}
The on-site distribution depends  on the random fields at the site $i$ and at the neighboring sites $j \in {\partial i}$, which we collectively denote by $\vec{\epsilon}_{ i}$.
 The distribution of local fields sampled aver all sites reads
{\medmuskip=0mu
\thinmuskip=0mu
\thickmuskip=0mu
\begin{equation}
\label{PT}
P(h) = \frac{1}{N}\sum_{i=1}^N \delta(h - h_i^{\rm{eff}}) =   \frac{1}{N}\sum_{i=1}^N \delta \tonde{h - \epsilon_i  - \sum_{j \in \partial i} J_{ij} \sigma^z_j},
\end{equation}}
where the  $\{\sigma_j^z\}$ realize a typical classical configuration sampled from the Gibbs ensemble at  temperature $T\equiv \beta^{-1}$. This distribution 
in general depends on the temperature. However, here we restrict ourselves to temperatures $T\gg \sum_j  J^2_{ij}/W$, in such a way that we can neglect the polarizing influence of spin $i$ on its neighbors. 
We also assume statistical time-reversal symmetry, implying that the distribution of local fields $f(\epsilon)$ is even.  
In this case the probability of finding neighboring spins up or down, averaged over all those spins, is always
\begin{equation}
p(\sigma^z)=\int d \epsilon \frac{e^{-\beta \epsilon \sigma^z}}{2 \cosh (\beta \epsilon)} f(\epsilon) {= \frac{1}{2}},
\end{equation}
independent of the temperature. It follows that the distribution of local fields $P(h)$ in Eq.~(\ref{PT}) is $T$-independent as well. While it would not be difficult to take correlation effects at lower temperatures into account, our simplification eliminates a potential temperature dependence of the decay rates that arises trivially from a $T$-dependence of the local field distribution. It thereby allows us to focus on the effect of thermal fluctuations only. A case where correlation effects are strong and  $P(h)$ does depend significantly on temperature at low $T$ is instead analyzed in Ref.~\onlinecite{us_paper2}.

Notice that the above definition of the distribution $P(h)$ makes sense even without a thermal average and even if the system were completely frozen, realizing a single classical configuration sampled from the Gibbs distribution, but without invoking a dynamic averaging. This is because the average over the entire sample ensures that all possible local configurations are sampled and represented in the sum (\ref{PT}). This makes a thermal average superfluous, since $P(h)$ is self-averaging.

\subsection*{Locator approximation for the decay rate}
To characterize localization for the model \eqref{eq:HamCuevas}, we consider the decay rate of a local (spin-flip) excitation created at a site $0$ in the bulk of the lattice. In a localized system, the decay rate vanishes in the thermodynamic limit, even if one couples the system to a bath at the boundary. In the tree approximation in which loops are neglected, approximate expressions for the decay rate $\Gamma$ (via the boundary)  can be derived, based on the linearization \cite{abou1973selfconsistent,Feigelman:2010fy, IoffeMezard2010,semerjian2009exact,muller2013magnetoresistance,Yu2013} of recursion equations for the imaginary parts of Green's functions.  For Hamiltonians such as in Eq.~\eqref{eq:HamCuevas}, in appropriate units they take the form: 
\begin{equation}\label{eq:GenPolymer}
 \begin{split}
  \Gamma_R [\vec{\epsilon}, \vec{\sigma}_{\partial};\omega]  = \sum_{\mathscr{P}: 0 \to \mathcal{B}_R} \,   \prod_{s \in \mathscr{P}}\, \frac{J_\perp^2}{[\omega- h_s^{\text{eff}} \tonde{\epsilon_s, \vec{\sigma}_{\partial s}} ]^2},
 \end{split}
 \end{equation}
where the sum is over the exponentially many lattice paths $\mathscr{P}$ that connect  the bulk site $0$ to the boundary $\mathcal{B}_R$, assumed to be at lattice distance $R$ from  site $0$. By $\partial = \cup \partial_s$ we denote the collection of all spins that are neighbors of sites $s$ on any such path $\mathscr{P}$. 
Different paths in the sum contribute with amplitudes that are a product of \emph{locators}, one for each site. The denominators essentially correspond to the mismatch between the energy of the propagating excitation $\omega$ and the effective field at the site. 
This expression is obtained within the so-called forward scattering approximation, that corresponds to taking the leading order in the quantum fluctuations $J_\perp$. In particular, within this approximation, self-energy corrections to the denominators are neglected \cite{abou1973selfconsistent, muller2013magnetoresistance, pc2016forward}. If the lattice is locally tree-like, the spins $\vec{\sigma}_{\partial i}$ with which the excitation interacts are different from site to site, and therefore the locators at different sites are statistically independent. We also resort to this approximation when having in mind more general lattices which are at best locally tree-like. \footnote{Notice that in single particle problems the effect of self-energies can be accounted for approximately, by imposing a lower bound to the energy denominators of the locators as suggested already in Ref.~\cite{anderson1958absence}. This preserves the statistical independence of locators along the path. We implement these corrections in the calculations in Appendix \ref{app:Calculation}, the results of which are shown in Sec. \ref{sec:ManyPaths}.}

The excitation is considered to be localized whenever the typical value of $\Gamma_R$ decays exponentially in $R$ with a positive spatial decay constant $\gamma>0$, $\langle \log \Gamma_R \rangle = -R \gamma + o(R)$ (see the next section for a precise definition of the average).  The vanishing of this constant, $\gamma=0$, can thus be taken as signature for the onset of a delocalized phase.

\section{Fluctuation-enhanced decay rate}\label{sec:Averages}
The amplitude of each path in the sum \eqref{eq:GenPolymer} is affected by the interactions with the neighboring spins, as it depends explicitly on the variables $\vec{\sigma}_\partial$. We refer to these variables as the ``neighboring" or ``environmental" spins in the following.  It is natural to expect that different values of the decay rate are obtained depending on whether these variables are treated as \emph{frozen} in a typical configuration at inverse temperature $\beta$, or as \emph{liquid} dynamic variables that are allowed to fluctuate at each site and assume different configurations according to their thermal probabilities~\footnote{The difference between frozen and liquid environmental spins is particularly relevant for models where the local fields along a given path depend mostly on environmental spins off the path, rather than a fixed local random field. This is usually the case, for lattices with a large connectivity.}.
In the latter case, one can have rare fluctuations in the environmental spins that give rise to local fields that are more frequently close to $\omega$ than in a typical thermal configuration. In an optimized fluctuation of environmental spins, the abundance of small energy denominators is higher, which opens a more efficient decay channel for the excitation than a typical configuration \cite{cuevas2012level}. 
Since the probability for a small deviation of denominators from a typical thermal distribution only decreases with the square of the deviation, while its effect on the decay rate is linear, we generally expect that fluctuations enhance the decay. In other words, the annealed decay rate is expected to be strictly bigger than its quenched counterpart, except possibly at the localization transition. This will be confirmed by our explicit calculations below. 

{\subsection*{Delocalizing effects of coupling to a weak bath}
The above calculation of decay rates makes perfect sense in a system where the neighboring spins are non-dynamic, which is the case if the only terms in the Hamiltonian were Ising couplings, that preserve the $z$-component of the neighbors. At strong enough disorder this  constitutes a particular realization of a closed, many-body localized system. Now, we however want to extend the consideration of decay, or the conductance across a finite sample, to the case where the neighboring spins are weakly coupled to a bath, e.g., of phonons, which allow for a  slow flip rate of the neighbors, such that over sufficiently long times all possible neighbor configurations are sampled with their thermal weight. This amounts to a slow stochastic time evolution of the local fields seen on the paths of the lattice.

From previous investigations of time-evolving potentials, see e.g. Refs.~\cite{gopalakrishnan2015low, abanin2016theory}, it is clear that over time  resonances will be created at finite distances, during which excitations will be displaced to new sites, from where later occurring resonances may carry them further in a random fashion. Such processes result in a slow diffusion. The weak coupling to a bath can also allow for inelastic transitions where some energy is exchanged with the bath. So both types of processes invariably induce a  small, yet finite diffusivity in the system that strictly speaking destroys localization. However, in the limit where the bath coupling is weak and the thermal fluctuations are very slow, and provided that we consider finite spatial distances, the induced diffusive processes become subdominant compared with direct decay. The latter does not involve the bath coupling and its average rate is independent of the frequency of thermal fluctuations. Here we focus on this regime of decay over finite spatial distances, that are such that transport via multi-step resonant processes or inelastic processes involving energy exchange with the bath are negligible. That is, we assume very slow fluctuation rates  and an associated time scale $\tau_f$ that exceeds typical times for excitations to tunnel coherently across the finite distance we are considering. We then ask about two limiting regimes of transport: We consider a regime, where we average the conductance or decay rates over times $\gg \tau_f$, which we refer to as the liquid, since fluctuations of the effective disorder potential are sampled over that time. We distinguish it from the frozen conductance or decay rates, which are relevant if we average only over time windows shorter than $\tau_f$. }
  
\subsection*{Annealed (liquid) and quenched (frozen) decay}  
To investigate the effect of thermal fluctuations, we describe the liquid environment  by computing the annealed average (which is essentially equivalent to the time-average) of the decay rates associated to each path in \eqref{eq:GenPolymer}, obtaining:
{\medmuskip=0mu
\thinmuskip=0mu
\thickmuskip=0mu
\begin{equation}\label{eq:DecayLiquid}
  \overline{\Gamma}_R[\vec{\epsilon};\omega]  = \sum_{\mathscr{P}: 0 \to \mathcal{B}_R}    \int \prod_{s \in \mathscr{P}} dh_s \, P(h_s| \vec{\epsilon}_{ s})  \min \grafe{1, \prod_{s \in \mathscr{P}}\frac{J_\perp^2}{[\omega- h_s ]^2}},
  \end{equation}}
where the field distribution $P(h_s| \vec{\epsilon}_{ s})$ is averaged over the thermal distribution of configurations of neighboring spins. Note that we have to be cautious when averaging the decay rate: While within the insulating phase it typically decreases exponentially with the path length, it might happen that on paths with rare configurations, where small denominators are more frequent, the product of locators becomes exponentially large. This is obviously an unphysical artifact which arises from our forward approximation and its neglect of self-energy corrections. Those would introduce correlations between the local fields and in particular suppress the effect of small denominators, ensuring that the decay rate never grows exponentially with distance. Indeed, from physical considerations, the decay rate can at best become of order $O(1)$. In Eq.~(\ref{eq:DecayLiquid}) we have remedied this artifact of our approximation by introducing an upper cutoff of $1$ on the locator product.   

To obtain a meaningful decay rate, we still need to specify how to average over the random fields $\epsilon_i$ that enter the above calculation.
In the case of a liquid environment, the decay rate should first be averaged over the annealed environmental spin variables, as described above; to obtain the typical decay rate, the resulting  $\overline{\Gamma} $ should then be logarithmically averaged over the local fields $\epsilon_i$. 
Notice that in this case the decay rate depends on the local random fields $\epsilon_i$  via the energy denominators, as well as via the distribution $P(h_s| \vec{\epsilon}_{ s})$ in \eqref{Ps}, since the  Boltzmann weight of the neighboring spins $\sigma^z_j$, depends on the local fields $\epsilon_j$.  In contrast, when the environment is treated as frozen (anticipating a fully localized phase) the average of $\log \Gamma$ should be taken over local fields and the configuration of neighboring spins, since the latter are quenched during the decay time. The resulting distribution is then simply given by $P(h)$ in \eqref{PT}. This leads us to define the following two spatial decay constants $\gamma_{F,L}$ that characterize the decay rates in frozen and liquid environment: 
\begin{equation}
\begin{split}
\gamma_{F}&= -\min \left(\lim_{R \to \infty}\int \prod_i \, d h_i \,  P(h_i)\frac{\log \Gamma_R }{R},0 \right), \\
\gamma_{L}&=- \min \left(\lim_{R \to \infty}\int \prod_i \, d \epsilon_i   f(\epsilon_i) \frac{\log \overline \Gamma_R}{R},0\right), 
\end{split}
\end{equation}
where $i$ runs over all lattice sites, and the subscripts F and L stand for ``Frozen" and ``Liquid", respectively. The expression for $\Gamma_R$ entering in the definition of $\gamma_F$ is given by \eqref{eq:GenPolymer} with the notation $h_s^{\rm eff}\to h_i$, while $\overline \Gamma_R$ in the definition of $\gamma_L$ is given in \eqref{eq:DecayLiquid}. As usual, the convexity of the logarithm implies $\gamma_{L} \leq \gamma_F$. This is in line with the physical expectation that  a fluctuating environment of neighboring spins can increase the abundance of small denominators. 

\subsection*{Coexistence of frozen and liquid phases?}
As discussed above, delocalization happens when $\gamma=0$. From the fact that the annealed decay rate is always bigger or equal to its frozen counterpart, one might expect that at a given temperature and at fixed disorder strength $W$, the critical value of the transverse field, $J_{\perp}^{(c)}$, at which $\gamma=0$ depends on whether the environment fluctuates or not, and that it might be smaller for a fluctuating environment than for a frozen environment, $J_{\perp, L}^{(c)} < J_{\perp, F}^{(c)}$. 
If that were indeed so, it would suggest a regime of coexistence, or bistability, $J_{\perp, L}^{(c)} < J_\perp <J_{\perp, F}^{(c)}$, in which both assumptions, a frozen, localized phase or a liquid, delocalized phase are self-consistent (whereby in this Gedankenexperiment we assume that thermal fluctuations are  permitted not through an external phonon bath, but rather through the system constituting its own bath, being in a many-body delocalized phase). However, such a  scenario will be ruled out below. Indeed we will show that in fact $J_{\perp, L}^{(c)} = J_{\perp, F}^{(c)}$, since the annealed and frozen averages become equal at criticality.

\section{One dimension - A single decay path }\label{sec:FrozAnn}
To illustrate the phenomenology of annealed and frozen decay rates in the simplest possible framework we consider first the case in which only one decay channel is accessible to the excitation, meaning that the sums in \eqref{eq:GenPolymer} and \eqref{eq:DecayLiquid} reduce to a single path, see Fig.~\ref{fig:Pictorial} (top). We caution that in this simple 1d case the localization-delocalization transition predicted by $\gamma=0$ should be taken with a grain of salt, since it is well known that for single particle Anderson localization in 1d, there is never a genuine delocalization. In that case $\gamma=0$ only marks the crossover to the weak localization regime, where the localization length becomes large, even though it does not diverge due to the relevance of backscattering at long distances and times. For many-body problems however, the latter are usually too weak to enforce localization~\cite{ChalkerDeLuca, ChalkerDeLuca2}  and $\gamma=0$ can still be taken as a reasonable estimate for the transition. 
Despite its simplicity this example already exhibits the transition (within the insulating phase of a system in a fluctuating environment) between the non-resonating insulator  and the intermittent metal. The latter fluctuates into becoming metallic for a small fraction of the time. Technically, it is characterized by dominant environmental configurations that saturate the bound in Eq.~\eqref{eq:DecayLiquid}, while  in the non-resonating insulator  the dominant fluctuations still have exponentially small decay rates, so that the bound in Eq.~\eqref{eq:DecayLiquid} remains irrelevant.

{The possibility and relevance of resonant transmissions in disordered conduction problems has long been recognized in static transmission problems, starting with the work by Lifshitz and Kirpichenkov~\cite{LifshitzKirpichenkov}. It shows up in the tunneling through insulating junctions as resonant transmission peaks as a function of energy~\cite{LifshitzGredeskulPastur82, AzbelSoven, SakKramer}, as well as in the Josephson coupling through disordered insulators~\cite{AzlamazovFistul,Shaternik}. Resonant transmission can also occur in classical wave propagation in inhomogeneous media~\cite{Kivshar10}. However, to the best of our knowledge previous works on static problems have not studied whether resonant transmission occurs  essentially at all energies in a given interval of interest, or whether it remains confined to narrow resonance windows. From our calculation below we will see that rare fluctuations will generically drive the system metallic (independent of the precise excitation energy), and we identify the conditions when such fluctuations provide the dominating channel of decay or transmission.} 

{The analysis of static disorder was extended to include the possibility of inelastic processes where resonant transmission occurs together with a finite energy exchange with phonon degrees of freedom that render the disorder potential dynamic.~\cite{GlazmanShekhter}\cite{Sokolovskii}. The essential result of these studies was that resonant transmission is still possible, albeit with the resonant transmission being smeared out over a somewhat larger energy window, but with comparable integrated transmission. 
These results suggest that the total transmission or decay rate is still well estimated by neglecting such inelastic processes. In our model they would correspond to processes where neighboring spins flip and absorb part of the energy of the excitation. We neglect such processes also for the reason that they come with small matrix elements, as the coupling to individual neighbors is weak.}

\subsection*{Frozen {\it vs} annealed decay rate}

We start by  deriving explicit expressions for the spatial decay constants $\gamma_{F,L}$. We focus on frequencies in the middle of the range of local excitations, choosing $\omega=0$ for simplicity and dropping the dependence on $\omega$ from now on. The frozen constant $\gamma_F$, when positive, is readily computed as:
\begin{equation}\label{eq:GammaF}
\gamma_{F}= -   \int d h \,  P(h) \log \left| \frac{J_\perp}{h} \right|^2.
\end{equation}
Recall that the distribution of local fields $P(h)$ does not depend on temperature, and therefore $\gamma_F$  does neither. In contrast, the  liquid decay rate does depend on temperature.
To compute it, we re-write:
\begin{equation}\label{eq:DR}
\begin{split}
\overline\Gamma_R[\vec{\epsilon}]&= \int_{-\infty}^\infty d x \,\text{min}\grafe{ e^{- R x},1} \, \mathcal{P}_R(x| \vec{\epsilon}),
\end{split}
\end{equation}
where $\mathcal{P}_R(x| \vec{\epsilon})$ is the probability (over the thermal configurations of neighboring spins) of finding a path amplitude that decays with spatial decay constant $x$. Formally it equals to:
\begin{equation}\label{eq:PRR}
\mathcal{P}_R(x| \vec{\epsilon})= \int \prod_{s=1}^R dh_s \,P(h_s| \vec{\epsilon}_s)
 \delta \tonde{\frac{2}{R}\sum_{s=1}^R \log \left| \frac{J_\perp}{h_s} \right|+ x}.
\end{equation}
We show in (\ref{eq:Prob2}) below that when evaluated for a typical realization $\vec{\epsilon}_{\rm typ}$ of the quenched random fields on path sites and on their neighbors, this 
probability can be re-written as:
\begin{equation}
\mathcal{P}_R(x| \vec{\epsilon}_{\rm typ})= e^{- R \Sigma(x)+ o(R)}.
\end{equation}
With this one readily obtains the large $R$ limit of $\overline \Gamma$,
 \begin{equation}\label{eq:GammaTyp}
\overline \Gamma_R[\vec{\epsilon}_{\rm typ}] = \int_{-\infty}^\infty d x\, \text{min}\grafe{ e^{- Rx},1}  e^{- R \Sigma(x)+ o(R)},
 \end{equation}
via a saddle point calculation. 

To obtain the probability of the spatial decay rate we represent the constraint in \eqref{eq:PRR} by an  integral over  an auxiliary variable $\xi$, which leads to:
\begin{equation}\label{eq:Prob}
\mathcal{P}_R(x| \vec{\epsilon})= \frac{R}{2 \pi}\int_{-\infty}^{\infty} d \xi \, e^{ R [i \xi x + \tilde{\Phi}_R(i \xi, \vec{\epsilon})]} 
\end{equation}
with:
\begin{equation}\label{eq:PhiT}
 \tilde{\Phi}_R(z, \vec{\epsilon})= \frac{1}{R}\sum_{s=1}^R\log \tonde{\int dh_s \, P(h_s| \vec{\epsilon}_s) e^{2 z \log \left| \frac{J_\perp}{h_s} \right|}}.
\end{equation}

The probability \eqref{eq:Prob} depends on the field realization $\vec{\epsilon}$ and is in general not self-averaging, being exponentially small in $R$; however, the quantity \eqref{eq:PhiT} is intensive. Its typical value equals
\begin{equation}\label{eq:TypPhi}
\begin{split}
 &\overline{\Phi}(z)= \int \prod_i [d \epsilon_i f(\epsilon_i)] \tilde{\Phi}_R(z, \vec{\epsilon})=\\
&\int d \vec{\epsilon}_s p(\vec{\epsilon}_s )\, \log \quadre{\int dh_s \, P(h_s| \vec{\epsilon}_s) e^{2 z \log \left| \frac{J_\perp}{h_s} \right|}},
 \end{split}
\end{equation}
where $p(\vec{\epsilon}_s )=f(\epsilon_s)\, \prod_{u \in \partial s} f(\epsilon_u)$. Therefore the probability \eqref{eq:Prob} evaluated on typical realizations $\vec{\epsilon}_{\rm typ}$, after a rotation in the complex plane, can be written as
\begin{equation}\label{eq:Prob2}
\mathcal{P}_R(x| \vec{\epsilon}_{\rm typ})=  c_R\int_{-i \infty}^{i \infty} d\xi \, e^{ R [\xi x + \overline{\Phi}(\xi)]} \equiv e^{- R \Sigma(x)+ o(R)}
\end{equation}
where $c_R$ is a constant that is sub-exponential in $R$.
For large $R$, the leading order term $\Sigma(x)$ can be obtained from a saddle point calculation as:
\begin{equation}\label{eq:Sigma}
\Sigma(x)= - [\xi^*(x) x+  \overline{\Phi}(\xi^*(x))] 
\end{equation}
where $\xi^*(x)$ is defined as the inverse function of: 
\begin{equation}\label{eq:SaddleLambdaA}
\begin{split}
x(\xi)&=- \overline{\Phi}'(\xi)=
 -\int d \vec{\epsilon}_s p(\vec{\epsilon}_s )\, \frac{\left \langle   \log \left| \frac{J_\perp}{h}\right|^2 \left| \frac{J_\perp}{h}\right|^{2\xi} \right \rangle}{\left \langle \left| \frac{J_\perp}{h}\right|^{2\xi} \right\rangle},
\end{split}
\end{equation}
with the notation $\langle \dots \rangle= \int dh P(h| \vec{\epsilon}_s) (\dots)$. 
Within this framework the typical decay constant (as in a frozen thermal configuration of the environment) is recovered setting $\xi^*=0$, with which one indeed finds 
\begin{equation}
\label{gamma_F=xtyp}
x_{\rm typ}=x(\xi^*=0)=\gamma_F
\end{equation}
upon comparing \eqref{eq:SaddleLambdaA} and \eqref{eq:GammaF}. Further, from \eqref{eq:Sigma} and using the identity $\overline{\Phi}(0)=0$ (cf. Eq. \eqref{eq:TypPhi})  one confirms that $\Sigma(x_{\rm typ})=0$ for this decay rate, as it must be for the logarithm of the probability of typical configurations.

After this digression, let us now return to evaluating the annealed spatial decay rate in a fluctuating environment, which we expect to be smaller than the one corresponding to typical configurations of the environment.
The configurations of $\vec{\sigma}_{\partial_i}$ that dominate the annealed average are usually exponentially rare ($\Sigma>0$). 
As a matter of fact, implementing the constraint in \eqref{eq:DecayLiquid} we find:
{\medmuskip=0mu
\thinmuskip=0mu
\thickmuskip=0mu
\begin{equation}\label{eq:Split}
\begin{split}
 \overline\Gamma'_R[\vec{\epsilon}_{\rm typ}] &=
  \int_{-\infty}^0 d x e^{- R \Sigma(x)+ o(R)}+  \int_{0}^\infty d x  e^{- R [x + \Sigma(x)]+ o(R)}.
\end{split}
\end{equation}}
 Given that we restrict to the regime in which $x_{\rm typ}>0$,  the first integrand in \eqref{eq:Split} assumes its maximum at a  value of $x$ outside the integration domain, and thus the integral is dominated by the contribution from the boundary $x=0$. The second term is dominated by the point $x_{\rm SP}$ that solves the saddle-point equation $\Sigma' = -1$ or:
\begin{equation}
\xi^*(x_{\rm SP})=1,
\end{equation}
as follows from \eqref{eq:Sigma} and \eqref{eq:SaddleLambdaA}. 
The solution reads:
\begin{equation}\label{eq:SaddleLambdaB}
\begin{split}
x_{\rm SP}&=
 -\int d \vec{\epsilon}_s p(\vec{\epsilon}_s )\, \frac{\left \langle   \log \left| \frac{J_\perp}{h}\right|^2 \left| \frac{J_\perp}{h}\right|^{2} \right \rangle}{\left \langle \left| \frac{J_\perp}{h}\right|^{2} \right\rangle}.
\end{split}
\end{equation}
This value belongs to the integration domain as long as $x_{\rm SP}>0$. In this case 
 one finds $\overline \Gamma_R= \text{exp}\tonde{- \gamma_L^{\rm naive} R + o(R)}$ with 
 \begin{equation}
 \gamma_{L}^{\rm naive} = -\overline{\Phi}(1) =-\int d \vec{\epsilon}_s p(\vec{\epsilon}_s )\, \log \left\langle  \left| \frac{J_\perp}{h} \right|^2  \right\rangle.
\end{equation}
However, once $x_{\rm SP}$ saturates to zero, the boundary value $x=0$ dominates the second integral, implying $\overline \Gamma_R= \text{exp}\tonde{- \gamma_L^{\rm int.met} R + o(R)}$ with 
 \begin{equation}
 \label{gammaLSigma0}
 \gamma_{L}^{\rm int.met} = \Sigma(x=0).
\end{equation}
Therefore, if positive, the liquid decay constant reads
\begin{equation}\label{eq:LiqFin}
\gamma_L= \Theta(x_{\rm SP})  \gamma_{L}^{\rm naive} + \Theta(-x_{\rm SP})    \gamma_{L}^{\rm int.met} .
\end{equation}

\subsection*{The fluctuations-induced transition in the insulator }
 \begin{figure}[ht]
    \includegraphics[width=1\linewidth]{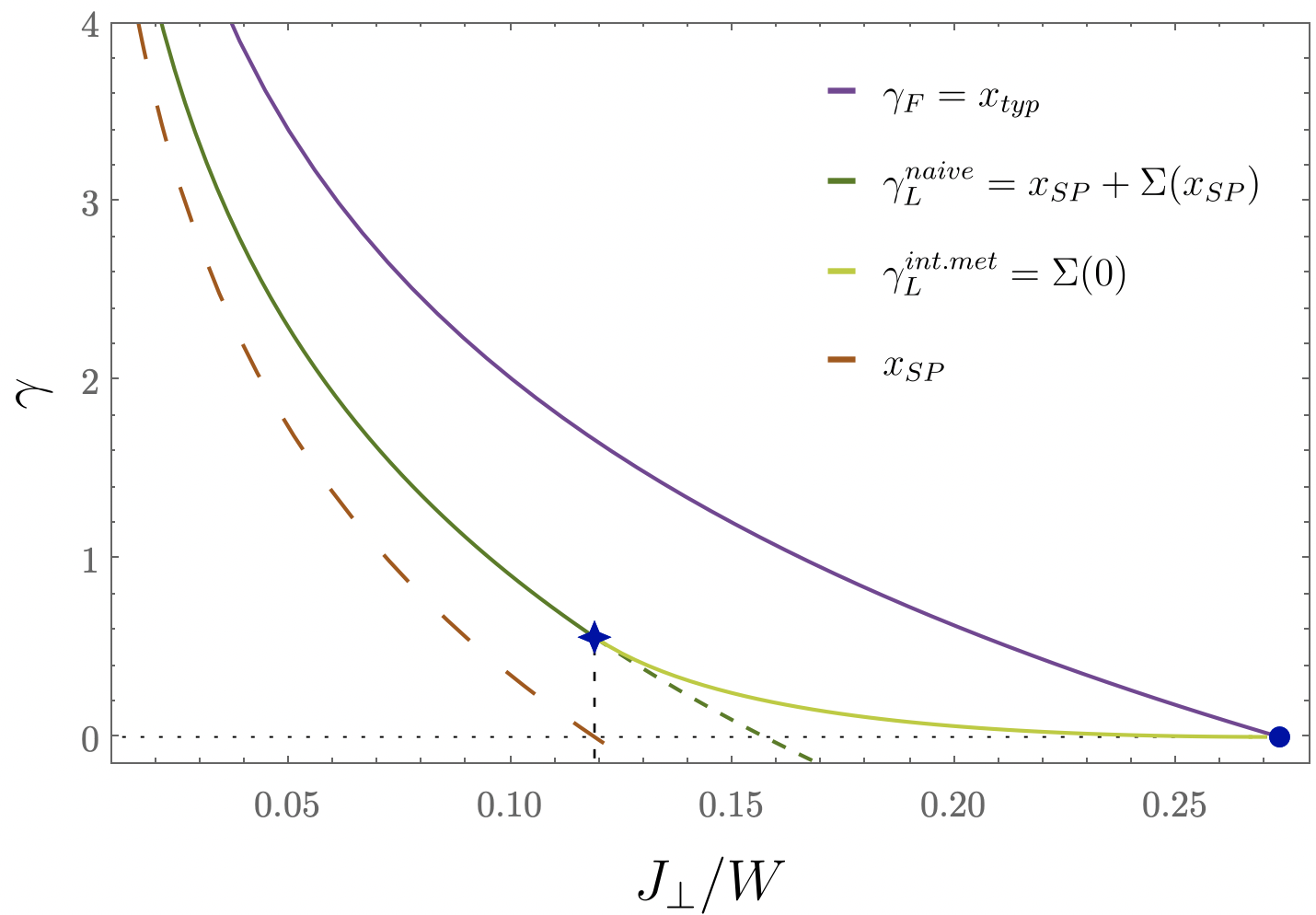}
    \caption{Comparison between the decay constants $\gamma_F$ (purple) and $\gamma_L$ (green) in frozen and liquid environments, respectively, for 
    { a chain (or a single path) each spin of which is coupled to  $N=2$ environmental spins}. Parameters are $W=1$, $\beta=0$ and $J_z=.26$. The blue star marks the transition between the \emph{non-resonating insulator} and the \emph{intermittent metal} of a system with a liquid environment. It occurs when the dominating fluctuations give rise to spatial decay constants $x_{\rm SP}=0$. In the non-resonating insulator the total liquid decay constant is larger than $x_{\rm SP}=0$ by an amount $\Sigma(x_{\rm SP})$, because the dominant fluctuations are exponentially rare temporal fluctuations. The blue dot marks the ``delocalization" crossover $J^{(c)}_\perp$ at which both decay constants become zero. At this point the system turns metallic, irrespective of the nature of the environment. The dashed green line shows the analytic continuation of the curve $x_{\rm SP}+ \Sigma(x_{\rm SP})$ into the region where the decay rate  of rare thermal fluctuations, naively evaluated, would take unphysical  negative values $x_{\rm SP}<0$. }\label{fig:FigPaths}
  \end{figure}

The behavior of the spatial decay constants $\gamma_F$ in \eqref{eq:GammaF} and $\gamma_L$ in \eqref{eq:LiqFin} is shown in Fig. \ref{fig:FigPaths} for a system with Hamiltonian \eqref{eq:HamCuevas} defined on a path with  $N=2$ environmental spins attached to each site as in Fig.~\ref{fig:Pictorial} (top). The fields $\epsilon_i$ are taken to be uniformly distributed in the interval $\quadre{-{W}/{2}, {W}/{2}}$, and the interactions $J_{ij} \equiv J_z$ are constant.  The analysis of this case of a single decay channel leads to the following conclusions:

\begin{itemize}
\item[(i)] \emph{Thermal fluctuations do affect spatial localization. The liquid decay rate is always smaller than the frozen one.} Indeed, for generic values of the parameters it holds that:
\begin{equation}
x_{\rm SP} + \Sigma(x_{\rm SP}) \leq x_{\rm typ},
\end{equation}
where $x_{\rm typ}$ is defined by $\Sigma(x_{\rm typ})=0$. Thus, $\gamma_L \leq \gamma_F$, consistent with the expectation that an annealed average over the environmental spins is in general dominated by configurations that are more resonant with the decaying excitation, but are too rare to contribute to the typical spatial rate $\gamma_F$ (as they have exponentially small probability, $\Sigma(x_{\rm SP})>0$). Fluctuations of the environmental spins thus lead to a larger effective localization length of local excitations.
 
\item[(ii)]  \emph{Within the insulator, a transition occurs in the annealed conductance. } The spatial decay constant $\gamma_L$ exhibits a non-analyticity when $x_{\rm SP}$ reaches zero. 
For weak hopping terms (or quantum fluctuations), $x_{\rm SP}>0$. This means that the thermal configurations that contribute dominantly to the decay rate give rise to a path weight {$\exp(-x_{\rm SP} R)$} that is itself exponentially small in $R$, and resonances are strongly suppressed. As the quantum fluctuations increase, the dominating spin configurations along a path become such that the far distance tunneling amplitude reaches values $O(1)$, which does not decay exponentially with $R$. This is the intermittent metal regime. However, the system still remains an insulator since the rate $\Gamma$ is now proportional to the probability of occurrence of such configurations through fluctuations, which is exponentially small in the path length and given by $\sim e^{- R  \Sigma(0)}$ with $\Sigma(0)= \overline{\Phi}(\xi^*(x_{\rm SP}=0)) $.

\item[(iii)] \emph{Thermal fluctuations do not affect the transition point out of the insulator. } The vanishing of the decay constant, $\gamma=0$, occurs at exactly the same value of the parameters in both the frozen and the liquid case: {for $\gamma_L= {\rm min}_{x \geq 0} \grafe{x+ \Sigma(x)}$ to be zero, one indeed needs that both $x = 0$ and $\Sigma(x=0)=0$, cf. Eq.~(\ref{gammaLSigma0}). Since  $\Sigma(x)$ vanishes for $x=x_{\rm typ}$ (by definition), this implies together with Eq.~(\ref{gamma_F=xtyp}) that at these parameters $\gamma_F=x_{\rm typ}=0$ as well.} Exactly at the transition, the configurations of the environmental spins that dominate the average conductance and lead to path weight of amplitude $\sim 1$ become typical. Therefore they contribute to the quenched conductance and the two conductances become equal.
\end{itemize}

In conclusion, even though the fluctuations of neighboring spins can indeed open more favorable decay channels for the decaying excitations, they do not shift the boundary of the localized phase. We note that in order to reach this last conclusion it was crucial to  impose the physical bound prohibiting exponentially large, unphysical path amplitudes: Failing to do so, one would miss the transition in (ii) and  would erroneously conclude that the two spatial decay constants $\gamma_L$ and $\gamma_F$ vanish at different points in parameter space (see the dashed green line in Fig. \ref{fig:FigPaths}).

\section{Optimizing over many paths}\label{sec:ManyPaths}
We now extend these results to the case of an exponentially large number of paths as it is relevant for higher dimensions $d> 1$. In particular, we will confirm the persistence of the transition between a non-resonating insulator and  an intermittent metal in the presence of a liquid environment. However, in addition, we identify another transition \emph{within the intermittent metallic phase}, which is the analogue of the glass transition in the related classical problem of a directed polymer in a random medium: On the more insulating side of the intermittent metal, the dominant path that occasionally becomes metallic is fixed. However, upon approaching the metallic phase, there are exponentially many paths that occasionally become metallic through thermal fluctuations. In other words, the optimal paths itself starts to fluctuate in space. 

In order to simplify the analytical treatment, we consider a Bethe lattice of branching number $k$, where additionally each site $s$ has its own $N$
environment spins $\sigma_j^z$ with $j \in \partial s$, see Fig. \ref{fig:Pictorial} (bottom). Each of these spins 
interacts with $\sigma_s^z$ through a coupling $J_{sj}\equiv J_z/ \sqrt{N}$, and sits in a field $\epsilon_j$ drawn independently for every site from the distribution $f(\epsilon)$. The quenched fields $\epsilon_s$ along the path are also random and independent, with identical distribution $f(\epsilon)$. We assume $N$ to be very large, so that the field transmitted to a site $s$ by the environmental spins,
\begin{equation}
 h^{\rm env}_s=\frac{J_z}{\sqrt{N}} \sum_{j \in \partial s} \sigma^z_j,
 \end{equation}
is a fluctuating Gaussian variable. The $\sigma_j^z$
are thermally fluctuating and have the probability
distribution $p(\sigma_j^z)=(1+ m_j \sigma_j^z)/2$ where $m_j = \tanh (\beta \epsilon_j )$.
Thus the $h^{\rm env}_s$ have means $M_s$ and variances $V_s$ (with respect to the thermal fluctuations) that depend explicitly on the random fields $\vec{\epsilon}_{\partial s}$, and read:
\begin{equation}
\begin{split}
M_s&\equiv M(\vec{\epsilon}_{\partial s})  = \overline{h^{\rm env}_s}=\frac{J_z}{\sqrt{N}} \sum_{j =1}^N m_j,\\
V_s& \equiv V(\vec{\epsilon}_{\partial s})  =  \overline{(h^{\rm env}_s)^2}-  \overline{h^{\rm env}_s}^2=\frac{J^2_z}{N} \sum_{j =1}^N (1-m_j^2),
\end{split}
\end{equation}
where the line denotes the average over thermal fluctuations of the $\sigma_j^z$ in their fixed random fields. Both $M_s$ and $V_s$ are random variables; however, $V_s$ has negligible fluctuations and tends to the fixed value
\begin{equation}
V_s \to V_\beta \equiv J^2_z \int d\epsilon \, f(\epsilon) [1- \tanh^2(\beta \epsilon)].
\end{equation}
In contrast, the mean $M_s$ fluctuates {from site to site} according to a Gaussian distribution with zero mean and variance:
\begin{equation}
v_M= J^2_z \int d\epsilon f(\epsilon) \tanh^2(\beta \epsilon)=J^2_z-V_\beta.
\end{equation}
Given a fixed configuration $\vec{\epsilon}_s=(\epsilon_s,\vec{\epsilon}_{\partial s})$ of the quenched local fields on the given sites $s$ and of the $N$ neighboring ones $j \in \partial s$, 
 the probability 
 (over the thermal fluctuations) to find
 an effective local field $h_s$ at the site $s$ is therefore a Gaussian that depends on $\epsilon_s$ and $\vec{\epsilon}_{\partial s}$ only through the combination:
 \begin{equation}\label{eq:MathEp}
 \mathcal{E}_s(\vec{\epsilon}_s)= \epsilon_s + M(\vec{\epsilon}_{\partial s}).
 \end{equation}
reading,
  \begin{equation}\label{eq:GaussM2}
 P(h_s|\mathcal{E}_s)= \frac{\text{exp} \tonde{-\frac{[h_s- \mathcal{E}_s]^2}{2 V_\beta}}}{\sqrt{2 \pi V_\beta}}.
 \end{equation}
 This corresponds to \eqref{Ps} and where
 we replaced the local variance with the averaged one. 
 To further simplify the calculations, we assume the distribution $f(\epsilon)$ to be Gaussian as well, with standard deviation $W$,
 \begin{equation}
 f(\epsilon)=\frac{e^{-\frac{\epsilon^2}{2 W^2}}}{\sqrt{2 \pi W^2}}.
 \end{equation}
 Then the distribution of \eqref{eq:MathEp}, 
 \begin{equation}
\begin{split}
\rho(\mathcal{E}_s)= \int d\epsilon_s \, f(\epsilon_s) \frac{e^{-\frac{(\mathcal{E}_s-\epsilon_s)^2}{2 v_M}}}{\sqrt{2 \pi v_M}},
 \end{split}
\end{equation}
 is itself Gaussian, and:
  \begin{equation}\label{eq:GaussM3}
 P(h_s)= \int d \mathcal{E}_s \, \rho(\mathcal{E}_s)\,P(h_s|\mathcal{E}_s)= \frac{\text{exp} \tonde{-\frac{h_s^2}{2 (J_z^2 + W^2) }}}{\sqrt{2 \pi (J_z^2 + W^2)}}
 \end{equation}
is independent of temperature as it has to be, see the discussion around Eq. \eqref{PT}. 
 
\subsection*{Biased distribution of quenched fields  along the optimal path}
Let us now discuss how to account for the presence of multiple paths. The main difficulty in this case
consists in simultaneously taking into account  the constraint on the liquid decay constant being non-negative, \eqref{eq:DecayLiquid}, and the statistics of the quenched fields $\vec{\epsilon}$. As an exponentially large number of
decay channels is available to the system, some of them have highly atypical 
distributions of quenched fields $\vec{\epsilon}$ along the path and its environmental sites. Those can lead to large deviations of path amplitudes. 
If the fluctuations of the local fields are sufficiently strong, these large deviations  affect the statistical behavior of sums of the form \eqref{eq:GenPolymer} and \eqref{eq:DecayLiquid}, in that they become dominated by a single (or at best a few) optimal path.

Let us first discuss the non-interacting limit of the problem, where the environmental spins are absent. In this case it is known that in the whole localized phase sums of the form \eqref{eq:GenPolymer} are dominated by one single {(or at best a few)} contribution(s) or ``optimal path(s)" \cite{abou1973selfconsistent, anderson1958absence,  aizenman2013resonant}, associated to a particularly large amplitude. In the non-interacting setting, on a Bethe lattice, this can be rationalized straightforwardly via a mapping to a directed polymer, by identifying the sum $\Gamma$ with the partition function of the polymer anchored at the origin of the lattice \cite{derrida1988polymers}. This mapping allows one to compute the spatial decay constant $\gamma$ as the free energy density of the polymer. Under this mapping the domination of $\Gamma$ by one {(or few)} optimal path(s) is the equivalent of the frozen (or glassy) phase of the polymer: the localized phase for a non-interacting problem always maps to the glassy phase of the corresponding polymer problem. Delocalization occurs when the amplitude of the optimal path becomes of order $O(1)$, which means  that exactly at the transition, among the exponentially many paths one typically finds just \emph{one} (or very few) that allow a decay to the boundary. \footnote{There is no transition in the  equivalent directed polymer problem, where the delocalization point merely maps to a point where the free energy density (the analogue of $\gamma$) vanishes.} Thanks to this mapping, the frozen path nature of the localized phase is straightforward to see, though one has to remember that the mapping relies on the forward approximation \cite{pc2016forward}.  However, the frozen path nature remains robust when self-energy corrections are added, as follows from rigorous results \cite{aizenman2013resonant, aizenman2011extended}. 
{In contrast, an interacting system with a fluctuating environment does not necessarily have to be in a frozen path phase, as we will see below.}

As we recall in Appendix \ref{app:Polymer}, the expression for the non-interacting decay constant $\gamma$ on a Bethe lattice, as derived in \cite{derrida1988polymers}, can be found from a convenient variational formulation \cite{Yu2013}: the amplitude of the dominating path is distributed like the maximum among $k^R$  path amplitudes with negligible mutual correlations, which reflects that the correlations among different paths on a Bethe lattice do not matter for the problem at hand.   
The dominating path will host a rather biased distribution of local fields, $\rho^{\rm opt}_k(\epsilon)$. The optimal distribution $\rho^{\rm opt}_k(\epsilon)$ along the optimal path can be determined by maximizing the expression of the decay constant, while making sure that the probability of observing such a distribution on a randomly picked path is at least $O(k^{-R})$, which ensures that such a path indeed typically occurs on a Cayley tree of depth $R$.
The logarithm of the probability $\mathbb{P}_\rho$ of finding a path of length $R$ with such a biased distribution is measured by the relative decrease in entropy, or Kullback-Leibler divergence, of $ \rho^{\rm opt}_k(\epsilon)$ as compared to the typical, unbiased distribution $\rho(\epsilon)$ of local quenched fields:
\begin{equation}\label{eq:Kullback}
\log \mathbb{P}_\rho = -R \int \,  \rho^{\rm opt}_k(\epsilon) \log \tonde{\frac{\rho^{\rm opt}_k(\epsilon)}{\rho(\epsilon)}} \, d \epsilon.\\
\end{equation}
Therefore:
\begin{equation}
\gamma=\min_{\rho^{\rm opt}_k : \frac{\log \mathbb{P}_\rho}{R} = -\log(k)} \grafe{ \int   \log \left| \frac{J}{\epsilon}\right|^2  \rho^{\rm opt}_k(\epsilon) d \epsilon},
\end{equation}
see Eqs. \ref{simplegamma} and \eqref{eq:Functional} in Appendix \ref{app:Polymer}.

This reasoning can straightforwardly be extended to the computation of the frozen decay constant $\gamma_F$ in the presence of environmental spins. Moreover, it will also allow us to derive compact expressions for the liquid decay constant $\gamma_L$, and to account for the bound on the path amplitude as we did in the case of a single path.

\subsection*{Directed polymers in frozen vs liquid environments}
{\it Frozen environment - } The decay rate in a frozen environment can be obtained following the recipe given in the previous section, which gives the correct result in a frozen  phase with essentially one optimal path. More generally it can be obtained by exploiting the directed polymer analogy, which we recall in Appendix \ref{app:Polymer}. The decay rate is obtained from the analog of the ``replicated free energy":
\begin{equation}\label{eq:PolimeroFrozen}
\begin{split}
\psi_{F}(\eta)&=- \frac{1}{\eta} \log \quadre{k \int dh \, P(h)  \Big| \frac{J_\perp}{h}\Big|^{2 \eta}},
\end{split}
\end{equation}
with the distribution of fields $P(h)$ as defined in \eqref{eq:GaussM3}. 
The replicated free energy $\psi_{F}(\eta)$ should be maximized over $0\leq \eta \leq 1$. More precisely, denoting  $\eta^*$ the argument which maximizes $\psi_F$ \eqref{eq:PolimeroFrozen}, one obtains for the decay constant
\begin{equation}
\gamma_F= 
\begin{cases}
\psi_{F}(\eta^*) &\text{    if    }\eta^* \leq 1\\
\psi_{F}(1) &\text{    if    }\eta^* > 1.
\end{cases}
\end{equation}
The first case applies in the frozen phase where the sum over paths is dominated by few contributions (sub-exponentially many in number). 
We discuss the detailed calculation of this function in Appendix \ref{app:Calculation}.  Here, we simply remark that in order for \eqref{eq:PolimeroFrozen} to be defined for $\eta \geq 1/2$, one needs to regularize the divergence arising due to small field denominators. Physically this is necessary since small denominators need to be resummed, which effectively cuts of the effect of too small denominators. In practice it suffices to impose a  cutoff to the integration, which regularizes the singularity from small denominators $h\approx 0$ \cite{anderson1958absence}, see Appendix \ref{app:Calculation} for  details. 
For a frozen environment we find that $\eta^* <1$ in the whole localized phase, exactly as in a non-interacting problem. This is expected, since the interactions enter the formalism only via the  distribution of the effective local fields.\\

We point out that it follows from the temperature independence of $P(h)$ (cf. Eq. \eqref{eq:GaussM3}) that the decay constant $\gamma_F$ is independent of $T$. This implies that also the location where it vanishes, i.e., the delocalization transition in a frozen environment, is temperature independent. Since we will show later on that the delocalization transition is independent of whether or not fluctuations are included, we reach the non-trivial conclusion that the transition to the metallic regime is entirely temperature independent in the model we consider here.\\

{\it Liquid environment - } Let us now turn to the more complex calculation of the decay constant in a liquid, fluctuating environment. Let us first assume that the sum \eqref{eq:GenPolymer} is dominated by an optimal path, with a configuration of fields $\vec{\epsilon}_s$ and thus of effective on-site fields $\mathcal{E}_s$ along the path. {We call $\rho_k^{\rm opt}(\mathcal{E}_s)$ the probability density describing the frequency with which the effective field $\mathcal{E}_s$ is encountered along that path, which of course differs from the distribution across the whole system,}
\begin{equation}\label{eq:BiasedDist}
\begin{split}
 \rho_k^{\rm opt}(\mathcal{E}_s) \neq \rho(\mathcal{E}_s)= \frac{e^{-\frac{\mathcal{E}_s^2}{2 (v_M+ W^2)}}}{\sqrt{2 \pi (v_M+W^2)}}. \end{split}
\end{equation}
The optimal path amplitude depends not only on the local fields $\epsilon_s$ at each site of the path, but via the fluctuating averages $M_s$ also on the fields on the neighboring sites $\partial s$. Let us denote by $\vec{\mathcal{E}}$ the collection of effective fields $\mathcal{E}_s$  along the optimal path.

We proceed by determining the decay constant $\gamma_L[ \rho_k^{\rm opt}]$ along this single optimal path as {described} in Sec.~\ref{sec:FrozAnn}, and subsequently determine the biased distribution \eqref{eq:BiasedDist} by optimizing the resulting constant over $ \rho_k^{\rm opt}$ under the constraint:
\begin{equation}\label{eq:KL}
\int \,  d \mathcal{E}\, \rho^{\rm opt}_k(\mathcal{E}) \log \tonde{\frac{\rho^{\rm opt}_k(\mathcal{E})}{\rho(\mathcal{E})}}  =\log k.
\end{equation}
The analogue of  \eqref{eq:PRR} describing the probability to encounter a thermal fluctuation giving rise to the decay constant $x$ along this path now takes the form: 
{\medmuskip=0mu
\thinmuskip=0mu
\thickmuskip=0mu
 \begin{equation}
 \begin{split}
 \mathcal{P}_R(x| \vec{\mathcal{E}}) &= \int \prod_{s} dh_s \,P(h_s| \mathcal{E}_s) 
 \delta \tonde{\frac{2}{R}\sum_{s=1}^R \log \left| \frac{J_\perp}{h_s} \right|+ x}.
 \end{split}
 \end{equation}}
 Let us denote with $\vec{\mathcal{E}}^{\rm opt}_{\rm typ}$ a realization of fields along the path, that is typical with respect to the unknown distribution \eqref{eq:BiasedDist}. 
 We have
  \begin{equation}
 \begin{split}
 \label{P_R}
 \mathcal{P}_R(x| \vec{\mathcal{E}}^{\rm opt}_{\rm typ}) &= e^{-R \Sigma[x; \rho_k^{\rm opt}] + o(R)},
 \end{split}
 \end{equation}
 where $\Sigma[x; \rho_k^{\rm opt}] $ is now a functional of the biased density $\rho_k^{\rm opt}$, defined as:
 \begin{equation}\label{eq:SigmaMF}
\Sigma[x; \rho_k^{\rm opt}]= - \tonde{\xi^*x+  \overline{\Phi}[\xi^*; \rho_k^{\rm opt}]}.
\end{equation}
Here 
 \begin{equation}
 \begin{split}
 \label{EqPhi}
&  \overline{\Phi}[\xi; \rho_k^{\rm opt}]=    \int d \mathcal{E} \,  \rho_k^{\rm opt}(\mathcal{E}) \log \quadre{\int dh  \,P(h| \mathcal{E})  \left| \frac{J_\perp}{h} \right|^{2 \xi} }\end{split},
 \end{equation}
 and $\xi^*=\xi^* [x; \rho_k^{\rm opt}]$ is obtained by inverting the saddle point condition $ x=  - {d \overline{\Phi}[\xi; \rho_k^{\rm opt}]}/{d \xi}$, which reads explicitly:
 \begin{equation}
 \label{xEQ}
 x= -\int d \mathcal{E} \, \frac{\rho_k^{\rm opt}(\mathcal{E})}{\omega_\xi(\mathcal{E})}\, \tonde{\int dh\, P(h| \mathcal{E})  \left| \frac{J_\perp}{h}\right|^{2 \xi} \log \left| \frac{J_\perp}{h}\right|^{2} }
 \end{equation}
with
\begin{equation}\label{eq:OmegaXi}
\omega_\xi(\mathcal{E})=\int dh  \,P(h| \mathcal{E})  \left| \frac{J_\perp}{h} \right|^{2 \xi}.
\end{equation} 
In this notation, we leave the dependence on the couplings $J_\perp$ and $V_\beta$ is implicit.  Similarly as in the frozen case, we use a cut-off around small fields to regularize the integral \eqref{eq:OmegaXi} for $\xi \geq 1/2$, see Appendix \eqref{app:Calculation} for details.
Following the same steps as in Sec. \ref{sec:FrozAnn} we obtain for the decay constant $\gamma_L$, defined via the decay 
$ \exp(-\gamma_L R ) \sim \int dx \min \left\{e^{-x R},1\right\} \mathcal{P}_R(x| \vec{\mathcal{E}}^{\rm opt}_{\rm typ})$:
 {\medmuskip=0mu
\thinmuskip=0mu
\thickmuskip=0mu
 \begin{equation}\label{eq:LiqFin2}
\gamma_L[ \rho_k^{\rm opt}]= \Theta(x_{\rm SP})  \gamma_{L}^{\rm naive}[ \rho_k^{\rm opt}] + \Theta(-x_{\rm SP}) \gamma_{L}^{\rm int.met}[ \rho_k^{\rm opt}] , \end{equation}}
where
\begin{equation}\label{eq:SaddleLambdaBmp}
\begin{split}
x_{\rm SP}&=
-\int d \mathcal{E} \, \frac{\rho_k^{\rm opt}(\mathcal{E})}{\omega_1(\mathcal{E})}\, \tonde{\int dh\, P(h| \mathcal{E})  \left| \frac{J_\perp}{h}\right|^{2} \log \left| \frac{J_\perp}{h}\right|^2 }
\end{split}
\end{equation}
and the two constants are given by
\begin{equation}
\label{eq:GammaNaive}
 \gamma_{L}^{\rm naive}[ \rho_k^{\rm opt}] ={-\overline{\Phi}[1; \rho_k^{\rm opt}]}=-\int d \mathcal{E} \, \rho_k^{\rm opt}(\mathcal{E}) \log [\omega_1(\mathcal{E})]
 \end{equation}
 and 
 \begin{equation}
 \label{boundgamma}
\gamma_{L}^{\rm int.met}[ \rho_k^{\rm opt}] =\Sigma[x=0;  \rho_k^{\rm opt}]=  -\overline{\Phi}[\xi^*; \rho_{k}^{\rm opt}].
\end{equation}
It is straightforward to check~\footnote{This follows from the fact that for fixed $\rho_{k}^{\rm opt}$, the function $x(\xi)$ defined in Eq.~\eqref{xEQ} is monotonically decreasing with $\xi$: $\partial x /\partial \xi \leq 0$, as can be verified with simple convexity arguments along the lines of those presented in Appendix \ref{app:NoMu1}. Therefore, if $x(\xi)$ is found to be negative at $\xi=1$, the equality $x=0$ will be attained at a value of $\xi<1$.}  from these two expressions that Eq.~(\ref{eq:LiqFin2}) can be compactly rewritten  as 
 \begin{equation}
 \gamma_L[ \rho_k^{\rm opt}]={\rm max}_{0\leq \xi\leq 1}\left\{-\overline{\Phi}[\xi; \rho_{k}^{\rm opt}]\right\}.
 \end{equation} 
These equations express the fact that the decay constant $\gamma_L$ corresponds to the naive annealed average $\gamma_{L}^{\rm naive}$, unless the expression for the latter is dominated by unphysical, exponentially growing contributions reflected by a negative growth rate $x_{\rm SP}<0$ at the saddle point.
In that case  the correct decay constant is given by \eqref{boundgamma}, which encodes the probability to encounter a thermal fluctuation that produces a metallic conduction along the path.
The logarithm of this probability is $\Sigma[x=0;  \rho_k^{\rm opt}]= -  \overline{\Phi}[\xi^*; \rho_{k}^{\rm opt}]$, cf. \eqref{eq:SigmaMF}, where $\xi^*$ is the solution of $-d\overline{\Phi}/d\xi=x=0$, or:
\begin{equation}\label{eq:KappaStar}
\int d \mathcal{E} \, \frac{\rho_k^{\rm opt}(\mathcal{E})}{\omega_{\xi^*}(\mathcal{E})}\, \tonde{\int dh\, P(h| \mathcal{E})  \left| \frac{J_\perp}{h}\right|^{2 \xi^*} \log \left| \frac{J_\perp}{h}\right|^{2} } =0.
\end{equation}

It remains to find the optimal biased density $\rho_k^{\rm opt}$. Our analysis below closely parallels the calculation in App.~\ref{app:Polymer} for the non-interacting case.
We optimize the decay rate subject to the Kullback-Leibler constraint \eqref{eq:KL} and the normalization constraint on $\rho_k^{\rm opt}$. 
To do so we define the functional:
\begin{equation}\label{eq:Func}
\begin{split}
&\mathcal{L}[\rho_k^{\rm opt}; \xi, \mu_1, \mu_2]=- \int d\mathcal{E} \,\rho_k^{\rm opt}(\mathcal{E}) \log[\omega_\xi(\mathcal{E})]+ \\
& \mu_1 \quadre{ \int d\mathcal{E}\, \rho_k^{\rm opt}(\mathcal{E}) \log \tonde{\frac{\rho^{\rm opt}_k(\mathcal{E})}{\rho(\mathcal{E})}}- \log k }+\\
& \mu_2 \quadre{ \int d\mathcal{E}\, \rho_k^{\rm opt}(\mathcal{E}) -1}.
\end{split}
\end{equation} 
The first line is $-\overline{\Phi}[\xi; \rho_{k}^{\rm opt}]$ with $\xi=\xi^*<1$ in the resonating phase or $\xi=1$ in the non-resonating phase, as can be seen from Eqs.~(\ref{EqPhi},\ref{eq:GammaNaive}, \ref{boundgamma}).
For any value of $\xi$, we find that the normalized solution of
\begin{equation}
\frac{\delta \mathcal{L}[\rho_k^{\rm opt};\xi, \mu_1, \mu_2]}{\delta \, \rho_k^{\rm opt}}=0
\end{equation}
reads
\begin{equation}\label{eq:OptimalDist}
\rho_k^{\rm opt}(\mathcal{E})= \frac{\rho(\mathcal{E}) \, [\omega_\xi(\mathcal{E})]^{\mu}}{\int \, d \mathcal{E} \rho(\mathcal{E}) \, [\omega_\xi(\mathcal{E})]^{\mu}},
\end{equation}
where we have substituted $\mu\equiv 1/\mu_1$. Injecting this form into the functional $\mathcal{L}$ we obtain the function
\begin{equation}\label{eq:FunctLiquiid}
\Psi_L(\mu, \xi)=-\frac{1}{\mu}\log \tonde{k \int \, d \mathcal{E} \rho(\mathcal{E}) \,  [\omega_\xi]^\mu},
\end{equation}
which still needs to be extremized with respect to $\mu$, {whereby it turns out that the relevant extremum is always a maximum.} {Notice that in this expression $\mu$ plays the same role as the parameter $\eta$ in the decay rate \eqref{eq:PolimeroFrozen} in a frozen environment. } Let $\mu^*$ the argument that maximizes $\Psi_L$, i.e., that solves:
\begin{equation}\label{eq:PolymerDerivative22}
\frac{\partial}{\partial \mu} \Psi_L(\mu, \xi)=0.
\end{equation}

Note that imposing the constraint \eqref{eq:KL} on the maximal Kullback-Leibler divergence only makes sense if the decay rate is maximized by a single path (or at most a sub-exponential number of paths). This is the case if and only if one finds  $\mu^*<1$. Otherwise one should impose no constraint, which is equivalent to setting $\mu=1$, in perfect analogy to the non-interacting case.
At the point where the solution of Eq.~\eqref{eq:PolymerDerivative22} reaches $\mu^*=1$, the quenched average over the random fields $\mathcal{E}$ becomes exactly equal to the annealed average, as one can see by setting $\mu=1$ in \eqref{eq:FunctLiquiid}. This corresponds to a melting transition in the associated polymer problem. 
While the Lagrange parameter $\mu$ controls the number of paths that contribute, we recall that the remaining parameter $\xi$ is associated with constraining the sampling of thermal fluctuations that lead to potentially negative decay constants. The expression for $\Psi_L$ still has to be maximized with respect to $\xi$ over the interval $0< \xi\leq 1$. We recall that a maximum with $\xi<1$ signals that the dominant thermal fluctuations turn the system temporarily metallic on the given path, indicating a rarely resonating insulator phase (intermittent metal).

The above formalism suggests the following algorithm to calculate the decay constant $\gamma_L$:
\begin{itemize}
\item[(a)] One first assumes $\xi=1$ and determines $\mu^*$ from 
\begin{equation}
\frac{\partial}{\partial \mu} \Psi_L(\mu, 1)\Big|_{\mu^*}=0.
\end{equation}
{The optimal distribution of $\mathcal E$ is then given by Eq.~\eqref{eq:OptimalDist} with $\mu=\mu^*$ and $\xi=1$. 
Upon substituting this in Eq.~\eqref{eq:SaddleLambdaBmp}, one then checks whether  the saddle point decay rate is positive, $x_{\rm SP}>0$, and thus physical. $x_{\rm SP}>0$ implies $\partial \Psi_L/\partial\xi > 1$, which guarantees that $\Psi_L(\mu^*,1)$ is indeed a maximum on the domain $0\leq \mu,\xi \leq 1$.}
When this holds true, it turns out that the maximizing $\mu^*$ always satisfies $\mu^* < 1$. This follows from a simple convexity argument given in Appendix \ref{app:NoMu1}. We recall that $\mu^*<1$ indicates that in this phase the decay rate is dominated by a sub-exponential number of paths. Since $\xi=1$ we further know that each of those contributes with a strictly exponentially decaying term ($x_{\rm SP}>0$). This regime thus corresponds to the deep, non-resonating insulator phase. The total spatial decay constant is obtained substituting the optimal distribution into \eqref{eq:GammaNaive} for the annealed thermal average along the optimal path, which yields  
\begin{equation}
\gamma_L= \gamma_L^{\rm naive}[\rho_k^{\rm opt}(\mu=\mu^*)]= \Psi_L(\mu^*, 1).
\end{equation}

\item[(b)] If the assumption $\xi=1$ and maximizing over $\mu$ leads to the physically inconsistent $x_{\rm SP}<0$ in \eqref{eq:SaddleLambdaBmp}, we know that $\Psi_L$ assumes its maximum for $\xi<1$. Assuming $\xi<1$ in turn means that the dominant thermal fluctuations are constrained in such a way that we make sure that the dominant path amplitudes just reach 1 (or $x_{\rm SP}=0$). In that case the physical bound on the path amplitude should be implemented by solving simultaneously the two saddle point equations:
\begin{equation}
\frac{\partial}{\partial \mu} \Psi_L(\mu, \xi)\Big|_{\mu^*, \xi^*}=0= \frac{\partial}{\partial \xi} \Psi_L(\mu, \xi)\Big|_{\mu^*, \xi^*},
\end{equation}
where the second equation 
ensures that $x=0$, see \eqref{eq:KappaStar}.
As long as the solution to both equations yields $\mu^* <1$, there is a single optimal path, which does not change under thermal fluctuations. Rare thermal fluctuations on this path turn it metallic, which dominates the decay. This regime is path-frozen.

Once the maximizing $\mu^*$ approaches $\mu^*=1$, the system undergoes a transition to a non-frozen phase in terms of the dominating decay paths. In this path-unfrozen regime, one has to set $\mu=1$, while $\xi^*$ is determined as the solution of $\frac{\partial}{\partial \xi} \Psi_L(1, \xi)\Big|_{\xi^*}=0$. 
One then obtains:
\begin{equation}
\gamma_L= \gamma_L^{\rm int.met}=
\begin{cases}
\Psi_L(\mu^*,  \xi^*) & \text{    if    } \mu^* \leq 1  \text{      \footnotesize{path-frozen}    }\\
\Psi_L(1,  \xi^*) & \text{    if    } \mu^* > 1\text{      \footnotesize{unfrozen}    }.
\end{cases}
\end{equation}
This yields the decay constant for the intermittent metal. 
\end{itemize}
It is straightforward to deduce from the above that this procedure is equivalent to the double maximization of $\Psi_L$ over a compact interval: 
\begin{equation}
\gamma_L = {\rm max}_{0\leq \mu,\xi\leq 1} \Psi_L(\mu,\xi ).
\end{equation}

Let us briefly discuss the resulting expressions for the spatial decay constants. Comparing \eqref{eq:FunctLiquiid} with the expression \eqref{eq:PolimeroFrozen} for the frozen case, we see that $\Psi_L(\mu, \xi)$ plays the role of the replicated free energy of a directed polymer, but now for a liquid environment. While in the frozen case the thermal realization of the environmental spins $\vec{\sigma}^z_{\partial }$ and the random local fields $\vec{\epsilon}$ are treated on the same footing, entering as quenched disorder into the distribution $P(h)$, in a liquid environment the thermal fluctuations are fast and averaged over first. This leads to the modified locator $ \omega_\xi$ in \eqref{eq:OmegaXi}, which takes the place of the simpler locator $(J_\perp/h)^2$ of the frozen problem. The averages over the random local fields and over the configuration of environmental spins, respectively, are associated with the two distinct parameters, $\mu$ and $\xi$. The parameter $\mu$ controls whether the number of  paths that dominate the decay rate is small (in the path-frozen regime, $\mu^*<1$) or exponentially large (in the path-molten insulator $\mu=1$).  The parameter $\xi$ instead is tuned such as to control the path amplitudes and prevent unphysical, exponentially growing contributions to the decay rate arising in the intermittent metal (which requires a non-trivial value $\xi^*<1$). 

These parameters are thus seen to take rather different roles. While at first sight it may look as if the transition from an intermittent metal to a non-resonating insulator, as signaled by the crossing of $\xi=1$, corresponds to some kind of glass transition, due to its formal resemblance to freezing transitions in replica theory, this is actually not the case. At a glass transition, the number of  metastable configurational valleys contributing significantly to the total free energy valleys shrinks from exponentially many to  $O(1)$. In our case, the role of configurational valleys of a directed polymer is assumed by distinct decay paths, and the corresponding unfreezing transition is indicated by $\mu^*$ reaching $1$. In contrast, $\xi^*$ becoming smaller than one indicates the vanishing of the dominant decay rate observed in rare thermal fluctuations, $x_{\rm SP}=0$, but it does not indicate the vanishing of the logarithm of the associated probability $\Sigma(x_{\rm SP})$ (which is what one might expect from a freezing transition). Indeed,  $\Sigma(x_{\rm SP})$ is still strictly positive at the transition to the resonating insulator.

 \begin{figure}[ht!]
    \includegraphics[width=.94\linewidth]{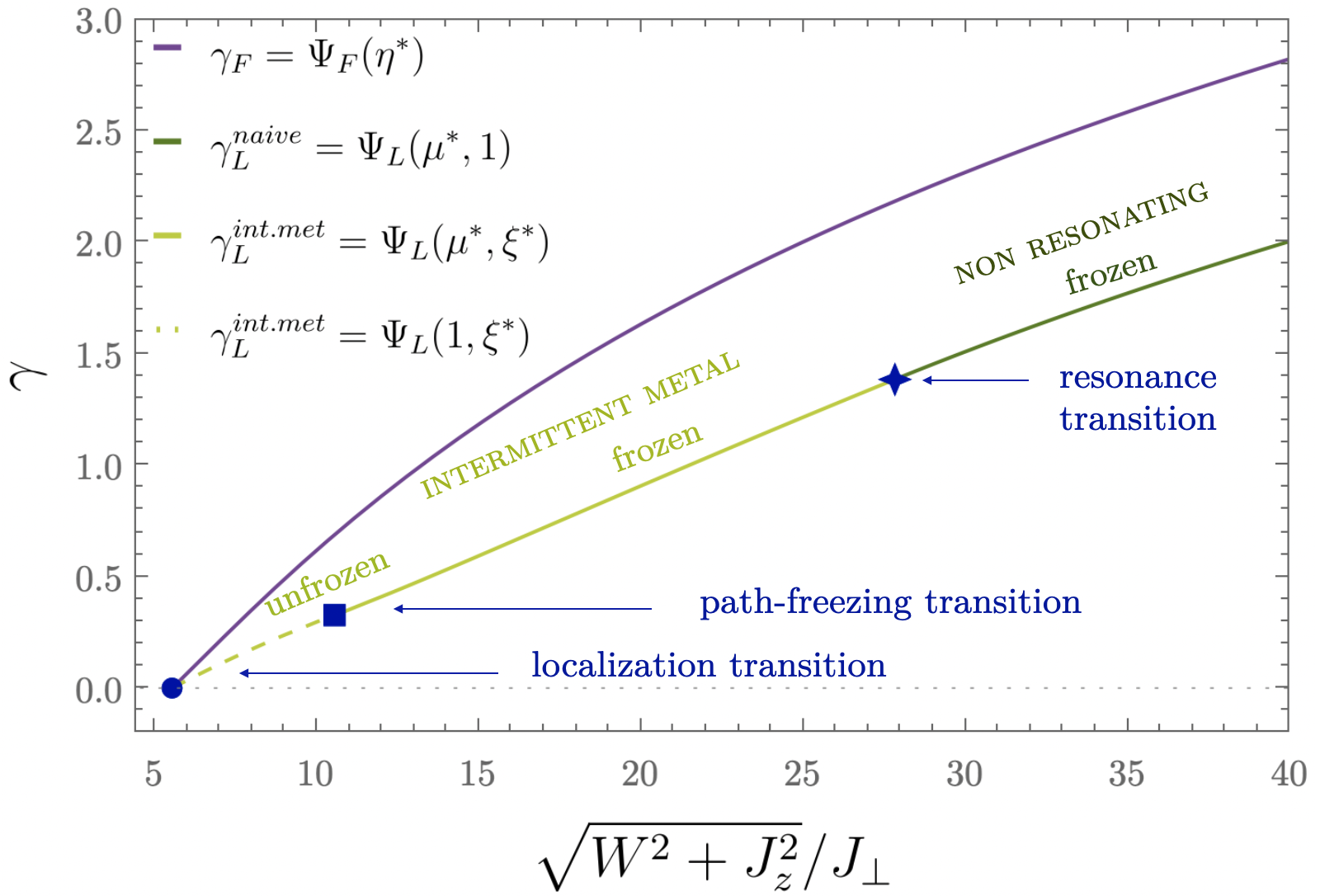}\\
    \vspace{.5 cm}
        \hspace{.25 cm}
        \includegraphics[width=.87\linewidth]{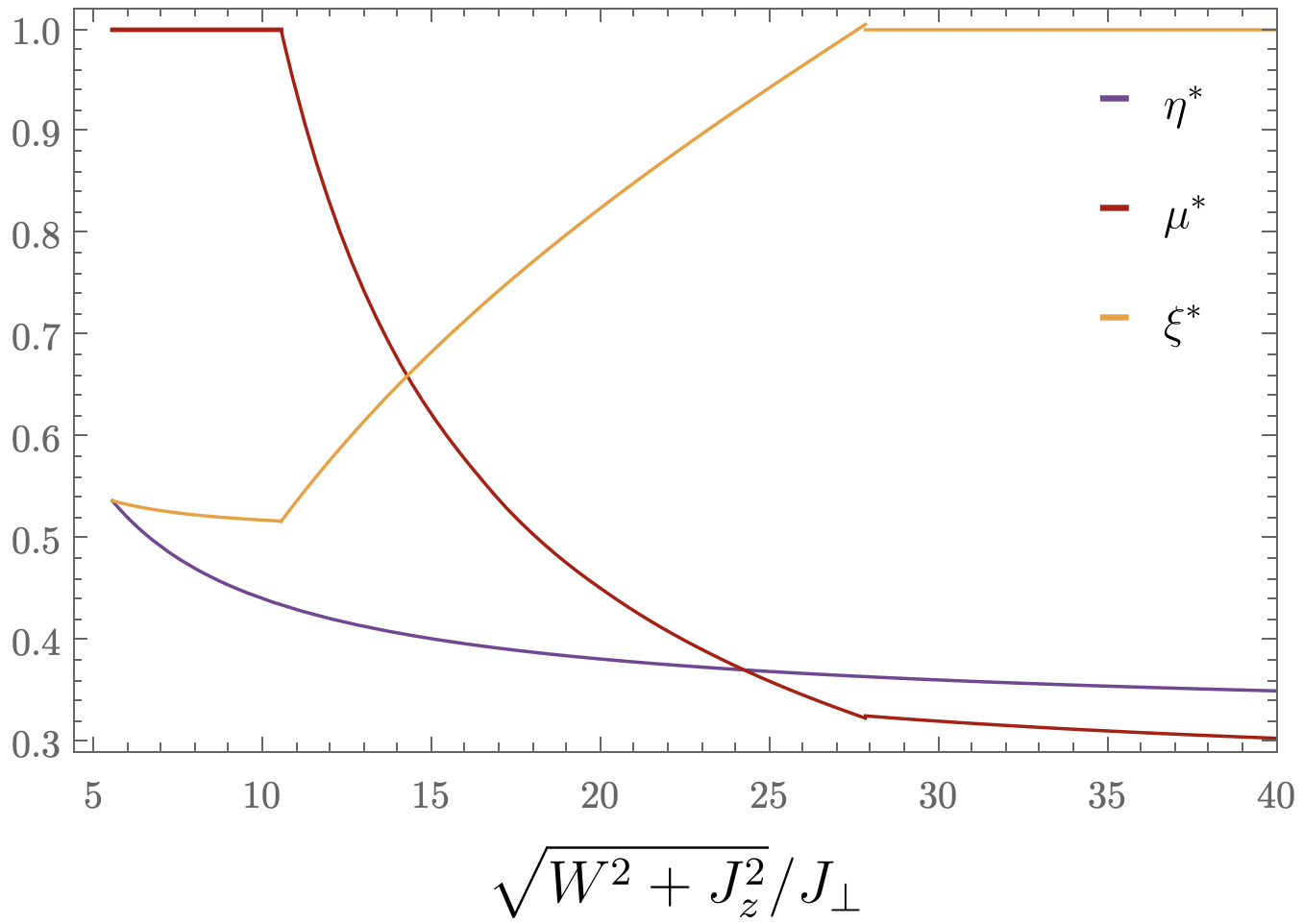}
    \caption{{\it Top. } Decay constants $\gamma_{F,L}$ at $T\to \infty$ in a frozen and liquid environment, respectively, on a Bethe lattice with branching number $k=2$, while every lattice spin couples to a separate environment consisting of $N\gg 1$ spins. The decay constants are plotted as a function of the ratio between the effective disorder strength $\sqrt{W^2+J_z^2}$, which controls the width of the distribution $P(h)$ of the local fields, and the amplitude of quantum fluctuations $J_\perp$. Parameters are $J_z=1= J_\perp$.   Localization is always stronger in a frozen environment, as seen by the strict inequality $\gamma_F> \gamma_L$ which holds all the way to the 
   delocalization transition at $\sqrt{W^2+J_z^2}/J_\perp=5.54$.
    Close to the metallic phase the insulator with liquid environment is in the intermittent metal phase. As the effective disorder increases, $\gamma_L$ undergoes first a path-freezing transition at $\sqrt{W^2+J_z^2}/J_\perp = 10.55$, and subsequently a transition to a non-resonating insulator at  $\sqrt{W^2+J_z^2}/J_\perp = 27.82.$ \\
    {\it Bottom. }  Evolution of the parameter $\eta^*$ that extremizes $\Psi_F(\eta)$ for a frozen environment, and of the parameters $\mu^*, \xi^*$ maximizing the functional $\Psi_L$ which captures the decay rate in a liquid environment. A non-resonating insulator is identified by $\xi^*=1$, while the path-unfrozen insulator has $\mu^*=1$. These two phases are always separated by an intermediate phase with non-trivial $\mu^*,\xi^*<1$ corresponding to an intermittently metallic, but path-frozen insulator. At the delocalization transition, $\xi^*=\eta^*$.}\label{fig:FigPathsMany}
  \end{figure}

\section{Three insulating phases brought about by thermal fluctuations}\label{sec:Discussion1}
In Fig. \ref{fig:FigPathsMany} ({\it Top}) we plot the decay constants $\gamma_F, \gamma_L$ for a Bethe lattice {with branching number $k=2$ (each site having $k+1=3$} neighbors on the lattice) as a function of disorder. {Fig. \ref{fig:FigPathsMany} ({\it Bottom}) shows the corresponding evolution of the liquid parameters $\mu, \xi$, as well as the parameter  {$\eta$ in the frozen case}, as defined in the previous section.} We refer to Appendix \ref{app:Calculation} for details about the computation. While a frozen environment only gives rise to a single insulating phase, the situation of a fluctuating environment is much richer. 
From the plots one can see that:
\begin{itemize}
\item[(i)] Upon decreasing the disorder or increasing the quantum fluctuations, the transition from a non-resonating insulator to an intermittent metal is preserved from the 1d situation discussed in Section~\ref{sec:FrozAnn}, even though now there are exponentially many paths  available for decay. This transition within the insulator probably comes closest to the thermal fluctuation-induced transition sought in Ref.~\cite{cuevas2012level}.
\item[(ii)] The liquid decay rate undergoes a further transition within the intermittent metal phase. It can be identified with an unfreezing transition of the corresponding directed polymer problem. Obviously, such a configurational unfreezing cannot occur in a 1d setting where only a single decay path is available. This transition between a path-frozen and an unfrozen regime always occurs within the intermittent metal regime, as we prove in Appendix \ref{app:NoMu1}. In contrast, in a system with a frozen environment the dominant decay always occurs along the same or the same few paths \cite{lemarie2019glassy}. It is the thermal fluctuations of environmental spins and the related changes in local fields that make the dominant decay paths of a system with liquid environment fluctuate. This path melting always takes place in a boundary regime adjacent to the transition to the metal. 
We will come back to the properties of this phase in Sec.~\ref{sec:chaos}.
\item[(iii)] The delocalization transition occurs at the same value of parameters, whether a frozen or a liquid environment are considered. This remains unchanged from the 1d case of a single path. The transition always occurs out of the path-unfrozen intermittent metal phase, where the parameter $\mu=1$ signals the contribution of exponentially many paths. This is proven in Appendix~\ref{app:NoMu1}. As we argued earlier, everywhere within the insulator the strict inequality $\gamma_F> \gamma_L$ holds.
\end{itemize}
We will discuss how the different insulating phases could  be distinguished by physical observables in Sec.~\ref{sec:phenomenology}.

\begin{figure}[ht!]
    \includegraphics[width=.99\linewidth]{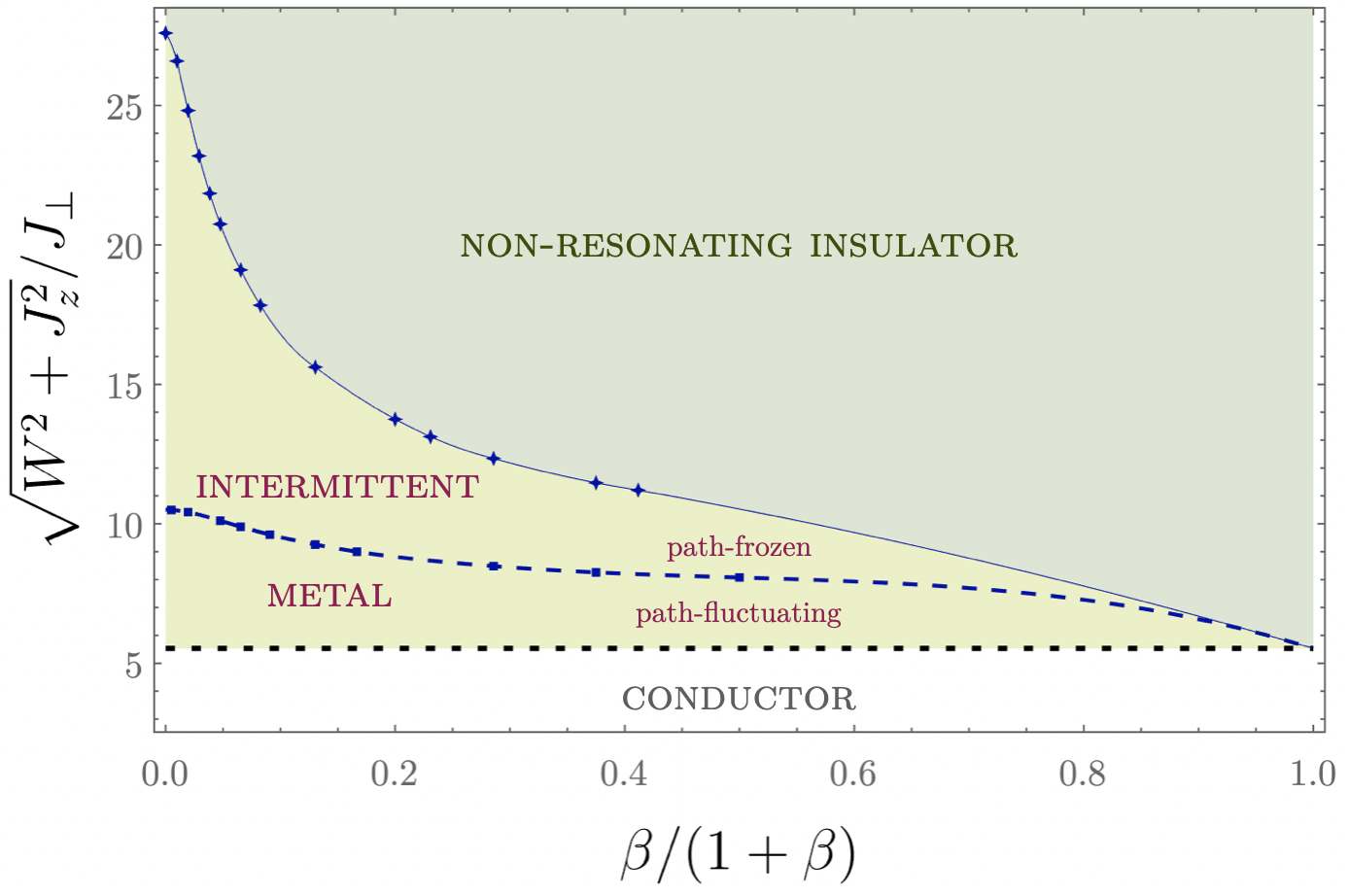}
    \caption{Temperature dependence of the critical values of the control parameter $\sqrt{W^2+J_z^2}/J_\perp$ separating the three distinct insulating phases in a fluctuating environment. The coupling to neighboring spins, $J_z$ is taken equal to the  spin hopping amplitude $J_\perp$, $J_z= J_\perp=1$. The dashed black line indicates the delocalization transition which is independent of temperature. The intermittent metallic phases (path frozen and fluctuating) shrink to zero in the limit $T\to 0$. Points are the results of the numerical calculation, while the lines are fitting curves.}\label{fig:BetaCrit}
 \end{figure}

The analysis leading to Fig. \ref{fig:FigPathsMany} has been carried out at infinite temperature which maximizes the effect of fluctuations of neighboring spins. We have focussed on a model where the temperature has the sole effect of controlling the strength of fluctuations of local fields, while it does not affect the global distribution of local fields $P(h)$ in Eq.~(\ref{PT}). We recall that, as a consequence, the critical point that separates the localized from the delocalized phase is independent of temperature, since it depends only on the global $P(h)$.  Indeed, the delocalization in frozen and liquid environments coincide, and  the transition in a frozen environment is a functional of the global $P(h)$ alone, as can be seen from Eq. \eqref{eq:PolimeroFrozen}.
 In contrast, the phase boundaries between the three insulating phases in a liquid environment are sensitive to the strength of thermal fluctuations, as  shown in Fig.~\ref{fig:BetaCrit}. The temperature dependence arises through the $T$-dependent width   of the distribution $P(h|\mathcal{E})$ which affects  the effective locators $\omega_\xi$ relevant in the fluctuating environment.   

The  intermittent metal phase shrinks with decreasing temperature. This is expected since the environmental fluctuations which promote those phases weaken as the temperature decreases. Both intermittent metallic phases disappear in the limit $T\to 0$. At the technical level this is reflected by the annealed calculation reducing to the quenched one.
 
We observe that temperature has a strong impact on the location of the resonance and path-freezing transitions. This is so even if we take the coupling to the neighbors $J_z$ to equal the spin hopping $J_\perp$, which is significantly smaller than the disorder $W$ needed for an insulator (cf. Fig. \ref{fig:BetaCrit}). One might thus expect temperature effects to be mild, as they shift local fields only by quantities of order $J_z$ as compared to the width of their global distribution $W$ .
However, the dominant locators on the dominating paths have still denominators  that are significantly smaller than $W$, and those react sensitively to shifts of the local field due to fluctuations in the thermal environment.
 \footnote{For instance, the resonance transition occurs when $x_{\rm SP}=0$, which from Eq.\eqref{eq:SaddleLambdaBmp} is seen to coincide with the vanishing of the double average:
 {\medmuskip=0mu
\thinmuskip=0mu
\thickmuskip=0mu
\begin{equation*}
\begin{split}
\int d \mathcal{E} \, {\rho_k^{\rm opt}(\mathcal{E})}    \mathbb{E}_{\mathcal{E}}\quadre{\log \left| \frac{J_\perp}{h}\right|^{2}}=0, \hspace{.25 cm}
 \mathbb{E}_{\mathcal{E}}\quadre{\otimes}= \frac{\int dh\, P(h| \mathcal{E})  \left| \frac{J_\perp}{h}\right|^{2} \otimes}{\int dh\, P(h| \mathcal{E})  \left| \frac{J_\perp}{h}\right|^{2}}.
\end{split}
\end{equation*}
}
{A decrease in temperature shrinks the width of both the distributions $P(h | \mathcal{E})$ and $\rho_k^{\rm opt}(\mathcal{E})$. The resulting shift of the effective  density of states are concentrated in a region where the logarithm has a strong dependence on its argument. Note  that the width of  $\rho_k^{\rm opt}(\mathcal{E})$  is indeed significantly smaller than $W$, as it is reduced by the weighting function $[\omega_\xi(\mathcal{E})]^{\mu}$, which strongly enhances small energies. Even moderate shifts of local fields (of order $J_z < W$) due to thermal fluctuations can thus sensitively affect the double average.}}

  \subsection*{No coexistence of frozen and liquid phases}
{Let us now address the question of the possibility of a coexistence of frozen and liquid phases in this class of models. With the above results at hand we can now show explicitly that this is excluded since the two decay rates $\gamma_F, \gamma_L$ vanish exactly at the same value of the hopping and disorder parameters. To prove this, since $\gamma_F \geq \gamma_L \geq 0$, it suffices to show that $\gamma_L = 0$ necessarily implies $\gamma_F = 0$. As we show in Appendix~\ref{app:NoMu1}, in the presence of a liquid environment the transition to delocalization always occurs inside a path-unfrozen phase, where the parameter $\mu$ takes the value $\mu =1$. Setting $\mu=1$ in \eqref{eq:FunctLiquiid} the liquid delocalization transition ($\gamma_L=0$) requires}
\begin{equation}
\begin{split}
\gamma_L&=-\log \quadre{k \int \, d \mathcal{E} \rho(\mathcal{E}) \,  \omega_{\xi^*}}\\
&=-\log \quadre{k \int \, dh \,P(h) \, \Big|\frac{J_\perp}{h}\Big|^{2 \xi^*}}=0,
\end{split}
\end{equation}
where {$P(h)$ is the unconstrained distribution defined in Eq.\eqref{eq:GaussM3}. }
Here $\xi^*$ is fixed by the condition: 
\begin{equation}\label{eq:CondPsi}
\int   dh\, P(h)  \log \left| \frac{J_\perp}{h}\right|^{2} \left| \frac{J_\perp}{h}\right|^{2 \xi^*}=0.
\end{equation}
Assuming $\xi^*>0$ we can divide each of these two vanishing quantities by some strictly positive quantities to obtain the equation: 
\begin{equation}
\frac{\log \quadre{k \int \, dh \,P(h) \, \Big|\frac{J_\perp}{h}\Big|^{2 \xi^*}}}{\xi^*}= \frac{\int   dh\, P(h)  \log \left| \frac{J_\perp}{h}\right|^{2} \left| \frac{J_\perp}{h}\right|^{2 \xi^*}}{\int   dh\, P(h) \, \left| \frac{J_\perp}{h}\right|^{2 \xi^*}},
\end{equation}
which is equivalent to the condition for a maximum of the  frozen decay functional $\Psi_F$,
\begin{equation}\label{eq:CondXiNu}
\eta^2\frac{d}{d \eta} \Psi_F(\eta)\Big|_{\eta=\xi^*}=0.
\end{equation}
This means that when the liquid decay vanishes, $\gamma_L=0$, the extremizers $\xi^*$ for the liquid case and $\mu^*$ for the frozen case coincide, $\xi^*=\eta^*$. This is seen in the explicit solution of Fig.~\ref{fig:FigPathsMany} ({\it Bottom}) . The frozen decay is given by  Eq.~\eqref{eq:PolimeroFrozen}. By virtue of Eq.~\eqref{eq:CondPsi} it vanishes as well, $\gamma_F=0$, which completes our proof. {We conclude that the sequence of transitions shown in Fig.~\ref{fig:FigPathsMany} and the coincidence of liquid and frozen delocalization hold for any choice of model parameters (such as the connectivity $k$, e.g.).} 
We recall that our  treatment of the decay is only approximate in that it is essentially a forward approximation where small denominators are suitably regularized. This captures
 only approximately the (anti)correlations between the locators at subsequent sites along the path that are induced by the self-energy corrections \cite{parisi2019anderson}. However, despite this approximation, we find that at the localization transition the values of $\eta^* =\xi^*$ come close to the exact value of $1/2$, which follows from a symmetry argument~\cite{abou1973selfconsistent}. 

The physical reason for the coincidence of the liquid and the frozen delocalization transitions, $\gamma_L=\gamma_F=0$ is relatively straightforward.
For any given thermal realization of the neighbor spin configuration, the local fields are as in a frozen configuration. For static configurations it is known that on a Cayley tree there is an optimal path that dominates the decay. The thermally fluctuating problem can have a non-exponential decay rate only, if a typical thermal realization of neighbor spin configurations gives rise to non-exponential decay, otherwise non-exponential decay would only occur in exponentially rare fluctuations. However, the requirement that typical configurations give rise to non-exponential decay is exactly the same as {that} for delocalization within a frozen environment.

\section{Phenomenology of the insulating phases}\label{sec:Discussion2}
\label{sec:phenomenology}

\subsection*{Fluctuating decay paths: path chaos}
\label{sec:chaos}
Our analysis in a liquid environment reveals the fact that thermal fluctuations of the neighbor spin configurations can favor different paths to become the instantaneously optimal decay channels. In other words, in the course of time the dominating decay path fluctuates, because the effective local field distribution fluctuates. However, this happens only close enough to the delocalization transition, in the phase that we dubbed the path-unfrozen phase.
This is a close analogue of a phenomenon observed in numerical studies of Anderson localization, where optimal decay paths were found to be very sensitive to changes in the disorder potential.~\cite{lemarie2019glassy} In the closely related classical problem of a directed polymer in random media, and more generally in spin glasses and similar glassy systems, a related kind of effect is known under the notion of ``chaos" \cite{McKayChaos,bray1987chaotic, CrisantiRizzoChaos}: The optimal (ground state) configuration of the glassy system is very sensitive to changes in the disorder realization, or to control parameters such as the temperature, in such a way that sudden jumps of the ground state configuration as a function of those parameters may occur, provided the system has enough time to equilibrate. 

The last point constitutes an important difference between quantum localization problems and classical glasses, however. The way these two classes of systems explore the available configuration space (i.e., the various paths for our localization problem) is fundamentally different. A glassy system usually has to overcome huge energy barriers in configuration space to settle into a new low energy configuration. This may take extremely long time, and thus often results in the system falling out of equilibrium, displaying aging, etc. This renders the observation of {\em equilibrium} chaos in glasses very challenging \cite{BeaChaos}.
The analogue of chaos in quantum localization problems instead is probed very differently. By its very nature, quantum mechanics probes all available decay paths simultaneously and instantaneously, like in a tunneling problem with parallel tunneling channels. In particular, the system usually retains no memory of  optimal paths that were favored by past disorder configurations that arise in the course of fluctuations. Insofar the decay channels react instantaneously to the fluctuations in the disorder realization and thus offer an interesting experimental route to studying chaos phenomena, which are much harder to probe in their thermodynamic analogues.

\subsection*{Experimental distinction of insulating phases}
 It is  interesting to discuss how the three insulating phases could be distinguished by experimental observables. 
 For electrons, or other excitations carrying a conserved charge,  the decay rates are related to the system's conductance across a mesoscopic sample, if direct tunneling across the sample is relevant.  In path-frozen phases the conductance is essentially dominated by a unique optimal path connecting the two leads. (In finite dimensions, other than on a Bethe lattice, such an optimal path is only defined up to small, spatially local fluctuations.)
 Accordingly, the conductance is strongly susceptible to disturbances that alter the local fields on that path. Such local perturbations could be produced, e.g., by an atomically sized tunneling tip in the close vicinity of the surface of a 2D sample. By scanning the sample surface laterally (parallel to the lead interfaces) one may expect a sudden spike or drop in conductance as the tip approaches the dominating conductance channel. In an un-frozen, path-fluctuating insulator instead, the dominant path is constantly fluctuating itself and any static local disturbance has little effect on the conductance. 
 
Local disturbances in the path-frozen phase of Anderson insulators can generally modify the localization properties of excitations, in case they happen to affect the optimal propagation channel of a relevant single particle wavefunction. It has recently been shown in single particle problems that indeed optimal decay paths can abruptly change as the potential landscape is altered~\cite{lemarie2019glassy}, and analogies to  closely related shocks and avalanches in glassy directed polymer problems have been drawn.
It might be interesting to study the continuously occurring ``wavefunction shocks" due to thermal fluctuations and the associated spatial range of paths that contribute to the conductance. 
 
Probing the transition between the intermittent metal and the non-resonating regimes is more subtle.  
Indeed, it requires to determine whether the exponentially rare  fluctuations that dominate the average (annealed) conductance correspond to a decay rate that is itself exponentially small or whether it is $O(1)$, meaning that the sample is intermittently  metallic.
The occurrence of a metallic-like fluctuation, even just over brief time windows, opens at least in principle the possibility for non-negligible, sample-spanning coherence effects in the conductance. In particular, Fabry-P\'erot-like oscillations of the conductance as a function of the  distance between two reflecting barriers inside the sample or at its boundaries, can occur with significant amplitude only if, at least for some instants of time, the transmission from barrier to barrier is quasi-metallic and multiply reflected waves can potentially interfere with each other.   
In contrast, more local quantum interference effects, such as weak localization due to  a sample-threading magnetic flux and the ensuing alteration of conductance would hardly allow to distinguish between a non-resonating insulator and an intermittent metal. Indeed, both regimes are effected by magnetic fields. In fact, strong insulators are usually disproportionately more affected than weak ones, since local interference effects are exponentially amplified in strong insulators. ~\cite{shklovskii1991hopping, syzranov2012strong, muller2013magnetoresistance}

The transition from non-resonating insulator to  intermittent metal also shows in a kink in the exponential decay of the conductance with sample length. Likewise one may expect 
 the conductance noise to bear a signature of the transition, since the conductance fluctuations in the resonating insulator start to be more strongly bounded from above. {In this context it would be interesting to revisit the reported growth and saturation of the Hooge parameter within the insulating phase~\cite{CohenOvadyahu}. }

\section{Conclusions}\label{sec:Conclusions}
We have analyzed the decay rate of local excitations interacting with thermally fluctuating environmental spins.  One of our driving questions was to understand whether internally generated fluctuations are able to shift the boundary of stability of the localized phase, a scenario that would entail the possibility of a coexistence regime between a delocalized and a (possibly metastable, but long-lived) localized phase in the vicinity of the localization-delocalization transition. It would also open the possibility to tune delocalization by temperature, via a mechanism that is different from the one usually discussed in the context of MBL~\cite{Basko:2006hh,gornyi2005interacting}, an idea originally raised in Ref.~\cite{cuevas2012level}.

Our analysis has indeed confirmed that the annealed rate of  decay (averaging over thermal fluctuations of the neighboring spins) is always bigger than the quenched decay rate (that of a typical, frozen environmental configuration) because the fluctuation average is dominated by 
 rare thermal configurations. However, the difference between the two rates diminishes upon approaching the metallic regime and disappears exactly at the delocalization transition. The location of that transition is therefore independent on whether the fluctuations are taken into account or not, and the putative coexistence or bistability scenario is ruled out. 

Nevertheless, we find that thermal fluctuations induce a rich phenomenology within the insulating phase. In general it hosts three different regimes, separated by two sharp transitions (at the level of our approximate description). All three regimes are characterized by an excitation decay rate  that decreases exponentially in the system size. At strongest disorder, the system is a \emph{non-resonating} insulator, where even exponentially rare, optimal fluctuations of the neighbor spin configurations give rise to exponentially weak decay. In other words, the system is insulating even in those rare moments in which the environment is particularly favorable and which therefore dominate the decay. In contrast, in the intermittent but path-frozen metal  the annealed time averaged decay rate is dominated by rare fluctuations of the neighboring spins that induce metallic-like behavior; nevertheless, their exponential rareness guarantees that the system is still insulating in the sense that the time averaged decay rate is exponentially small in the system size.  As in the non-resonating phase, the dominant decay path remains fixed and does not change, even though the environment fluctuates. This changes closer to the delocalization transition, where the analogue of a melting transition occurs and an intermittent metal with fluctuating optimal paths emerges. In this least insulating phase an exponentially large number of paths contribute to the time-averaged decay {(we recall that optimal paths are sharply defined in a Cayley tree approximation to real space, while in finite dimensional samples spatially local fluctuations around a dominant decay path always contribute as well, giving a finite width to the dominant channel)}. It is still true that at every instant of time, if one were to freeze the configuration of the environmental spins, one path would essentially dominate the decay. However,  this dominant path is now strongly sensitive to the thermally fluctuating effective local fields, in an analogous manner as ground states of glassy systems can change substantially with small modifications to the couplings or thermodynamic parameters. Here, we can prove for the model of a Cayley tree with neighboring spins that such a path-chaotic phase generically exists close to the delocalization transition.  As the latter is reached, the decay rate becomes order $O(1)$, i.e., independent of the system size. As mentioned above, its location does not depend on whether thermal fluctuations are accounted for or whether the environment is taken to be frozen.

{Let us compare our results to previous studies of similar static problems. It was found that the transmission through a wide barrier hosting hosting dilute, {\em randomly positioned}, but essentially identical impurities may be dominated by resonant transmission~\cite{lifshitz1979tunnel, lifshitz1982theory,rakih1987hopping}, depending on the energy of the propagating particle. This is analogous to our dichotomy between intermittent metal and non-resonating insulator, whereby the spatial average over the broad junction replaces our temporal average. However, we have considered a more general model, with random impurity potentials, which allows for a transition between non-resonant and rarely resonant insulator, even at transmission energies belonging to the support of the random Hamiltonian. In our case the transition is  tuned by the ratio of hopping and disorder strength, rather than by the energy. In contrast to static problems, our fluctuating setting entails the new possibility of a path-chaotic phase within the intermittent metal.}\\

We point out that despite some superficial similarities, the transition associated with the unfreezing  of the dominant path is different in nature from the crossover or putative transition, between a fully ergodic phase and non-ergodic delocalized phase, which has been controversially discussed, especially for non-interacting problems  on  Bethe lattices  or Cayley trees \cite{biroli2012difference, de2014anderson, monthus2011anderson, tikhonov2016anderson, sonner2017multifractality, kravtsov2018non, parisi2019anderson, biroli2020anomalous, HuseAltland2020,tarzia2020many}. 
Indeed, the path-unfreezing transition we have identified in this work occurs \emph{within} the insulating phase, even though it requires interactions with liquid-like fluctuating degrees of freedom. This transition entails a non-analytic behavior of the spatial decay constant $\gamma_L$ governing the decay rates of excitations in the thermodynamic limit (akin to localization lengths or Lyapunov exponents in non-interacting systems). In contrast, the putative transition between an ergodic and possibly non-ergodic delocalized phase, which was suggested to map the unfreezing of an associated effective polymer problem, takes place within the delocalized metallic phase where the participation ratio and fractal properties of non-localized wavefunctions evolve in a non-trivial manner. 

We recall that in the model discussed here the local fields in \eqref{eq:DecayLiquid} are sums of \emph{uncorrelated} contributions from environmental spins. However, in more realistic models correlations may establish among the environmental spins at low temperature. Those can induce a significant temperature dependence of the global distribution of fields, and thus affect the evolution of the insulating phases with temperature. If on top of that long range interactions play a crucial role, new phenomena such as pair-delocalization and spectral diffusion come into play~\cite{YaoDipolar, maksymov2020many, burin2006energy, smith2016many,rademaker2019bridging}.

\begin{acknowledgments}
V. Ros acknowledges funding by the ``Investissements d’Avenir” LabEx PALM (ANR-10-LABX-0039-PALM) and by the LabEx ENS-ICFP: ANR-10-LABX-0010/ANR-10-IDEX-0001-02 PSL*. She thanks the Paul Scherrer Institute in Villigen and the Galileo Galilei Institute (GGI) in Florence for hospitality and support during the completion of this work. M. M\"uller acknowledges funding from the Swiss National Science Foundation under Grant 200021\_166271 and thanks GGI Florence for hospitality.  
\end{acknowledgments}

\appendix
 \section{Directed polymer in random medium: a recap}\label{app:Polymer}
The free energy density of a directed polymer on the Bethe lattice has been computed explicitly in Refs. \onlinecite{derrida1988polymers,derrida1990directed}. On a finite Cayley tree of depth $R$, the partition function of the polymer is given by a sum over all paths $\mathscr{P}$ connecting the root $0$ to the boundary $\mathcal{B}_R$ at distance $R$, and it takes the generic form:
 \begin{equation}\label{eq:PartFunct}
\mathcal{Z}_{\beta, R}= \sum_{\mathscr{P}: 0 \to \mathcal{B}_R} \prod_{s \in \mathscr{P} } w_s,
 \end{equation}
where $w_s= e^{-\beta E_s}$ is the contribution of the site $s$, with $\beta$ the inverse temperature and $E_s$ a local energy. 
For a single-particle Anderson problem on the Bethe lattice with Hamiltonian
\begin{equation}
H= \sum_i \epsilon_i n_i - J_\perp \sum_{\langle i,j \rangle} \tonde{c^\dag_i c_j + c^\dag_j c_i},
\end{equation}
the decay rate $\Gamma_R$ computed in the forward approximation, 
\begin{equation}\label{eq:AppDec}
\Gamma_R=\sum_{\mathscr{P}: 0 \to \mathcal{B}_R} \prod_{s \in \mathscr{P} } \Big| \frac{J_\perp}{\epsilon_s}\Big|^2
\end{equation}
has exactly the form \eqref{eq:PartFunct} with $\beta=1$ and $E_s= -\log  |J_\perp/\epsilon_s|^2$,  where the $\epsilon_s$ are the random local potentials with distribution $f(\epsilon)$. We thus set $w_s= |J_\perp/\epsilon_s|^2$. The decay constant $\gamma$ can be identified with the free energy density of the polymer, 
 \begin{equation}\label{eq:QuenchedAv}
 \gamma= - \lim_{R \to \infty}\frac{\langle \log \Gamma_R \rangle }{R}.
 \end{equation}
 The exact results in Ref. \onlinecite{derrida1988polymers} show that $\gamma$ is fully determined by the function:
 \begin{equation}\label{eq:PsiNuSingle}
 \psi(\eta)=-\frac{1}{\eta} \log \quadre{k \int d\epsilon\, f(\epsilon) [w_s(\epsilon)]^{\eta}},
 \end{equation}
 where $k$ is the branching number of the Cayley tree. This can be recognized as the replicated free energy (per unit length) in a replica approach.
 
 Let us call $\eta^*$ the argument that maximizes $\psi$, satisfying
 \begin{equation}\label{eq:MinGen}
 \frac{d}{d \eta} \psi(\eta)\Big|_{\eta=\eta^*}=0.
 \end{equation}
It was shown  in Ref. \onlinecite{derrida1988polymers} that :
 \begin{equation}\label{eq:Gamma}
 \gamma= \begin{cases}
  \psi(\eta^*) & \text{  if   } \eta^* \leq 1\\
   \psi(1) & \text{  if   } \eta^* > 1,
 \end{cases}
 \end{equation}
 in agreement with the heuristic recipe that the replicated free energy should be maximized over the domain $0\leq \eta \leq 1$.
 The first regime corresponds to the polymer being in its frozen phase {(with broken replica symmetry)}, in which the partition function \eqref{eq:PartFunct} is dominated by a sub-exponential number of paths.  
 
As discussed in the main text, the spatial decay constant in a frozen environment can be obtained as a straightforward application of these identities, by replacing the distribution $f(\epsilon)$ with that of local fields, $P(h)$. In contrast, in the presence of a fluctuating environment these identities cannot be generalized straightforwardly, as one needs to enforce the physical constraint that the single path weights are bounded by $1$. 
In that case, it proves convenient to exploit a variational argument to determine the distribution of the quenched energy variables along the paths that dominate the decay rate. Here we recall this argument in the simpler setting of a directed polymer with no constraints, showing how it allows to recover \eqref{eq:Gamma}. 
 
Let us assume that \eqref{eq:AppDec} is dominated by a single path among the $k^R$ paths contributing to the sum. The local fields $\epsilon_s$ along the dominating, optimal path have an atypical statistics: they are not a typical sample from the distribution $f(\epsilon)$, but rather look like a sample from a biased distribution (to be determined) that we denote by $f^{\rm opt}(\epsilon)$. {The decay rate along this path is simply given by 
\begin{equation}
 \label{simplegamma}
 \gamma\equiv \gamma\{f^{\rm opt}(\epsilon)\}= -\int d\epsilon\, f^{\rm opt}(\epsilon)\, \log [{w}(\epsilon)].
 \end{equation}
 }
{However, to introduce some formal steps and functions whose more complex equivalents are used in the main text, we briefly derive it here by a formal detour.} 
Denoting by ${w}_s$ the weights along the optimal path, we can write the decay rate as:
\begin{equation}\label{eq:Uno}
\begin{split}
\Gamma_R&\approx \int dx e^{-R x} \delta \tonde{x+ \frac{1}{R}\sum_{s=1}^R \log {w}_s}\\
&=R \int dx e^{-R x} \int d\xi e^{i \xi\, R x+i \xi \sum_{s=1}^R \log {w}_s}.
\end{split}
\end{equation}
We introduce the function
\begin{equation}
\Phi(z)= \frac{1}{R} \sum_{s=1}^R \log {w}_s^{z},
\end{equation}
 and its average with respect to the biased distribution
 \begin{equation}
\overline{\Phi}(z)= \int d\epsilon\, f^{\rm opt}(\epsilon)\, \log [{w}(\epsilon)]^{z}.
\end{equation}
Taking the saddle point over $x$ in \eqref{eq:Uno} we obtain that $i \xi_{\rm SP}=1$, and $x$ drops out of the saddle point value. One then finds:
 \begin{equation}
 \label{simplegamma2}
 \gamma= -\lim_{R\to \infty} \frac{\log\Gamma_R}{R}=-\overline{\Phi}(1),
 \end{equation}
 which is equivalent to Eq.~ (\ref{simplegamma}).
 
In order to determine the unknown biased distribution $f^{\rm opt}(\epsilon)$, it suffices to maximize the functional Eq.~(\ref{simplegamma}) subject to the normalization constraint and the requirement that the {Kullback-Leibler divergence equal $\log(k)$}. That is, we should maximize
 \begin{equation}\label{eq:Functional}
 \begin{split}
 &\mathcal{F}[f^{\rm opt}, \mu_1, \mu_2]=- \int d\epsilon \, f^{\rm opt}(\epsilon) \log [{w}(\epsilon)]+ \\
 & \mu_1 \tonde{ \int d\epsilon \, f^{\rm opt}(\epsilon) \log \frac{f^{\rm opt}(\epsilon) }{f(\epsilon) }-\log k}+\\
 &\mu_2 \tonde{ \int d\epsilon \, f^{\rm opt}(\epsilon) -1},
 \end{split}
 \end{equation}
 where the Lagrange multiplier $\mu_1$ enforces that the optimal path occurs with a probability scaling as $\sim k^{-R}$, see \eqref{eq:Kullback} in the main text, and $\mu_2$ ensures the normalization of $f^{\rm opt}(\epsilon)$.
 The solution to 
 \begin{equation}
 \frac{\delta  \mathcal{F}[f^{\rm opt}, \mu_1, \mu_2]}{ \delta f^{\rm opt}}=0 = \frac{\partial  \mathcal{F}[f^{\rm opt}, \mu_1, \mu_2]}{ \partial 	\mu_2}
 \end{equation}
reads:
 \begin{equation}
 \label{ansatz}
 f^{\rm opt}(\epsilon)=\frac{f(\epsilon)  [{w}(\epsilon)]^{\frac{1}{\mu_1}}}{\int d\epsilon \, f(\epsilon)\, [{w}(\epsilon)]^{\frac{1}{\mu_1}}}.
 \end{equation}
 It remains to optimize with respect to $\mu_1$. Upon injecting the form of Eq.~(\ref{ansatz}) into \eqref{eq:Functional}  and changing notation to $\eta\equiv 1/\mu_1$, we find $\mathcal{F}(\mu_1= 1/\eta)= \psi(\eta)$, where $\psi(\eta)$ was given in \eqref{eq:PsiNuSingle}. We thus recover the exact result $\gamma= \psi(\eta^*)$, where $\eta^*$ satisfies \eqref{eq:MinGen}, under the assumption that one path dominates  the sum $\Gamma_R$.

However, as $\eta^* \to 1$, this assumption breaks down \cite{derrida1990directed} and the system leaves the frozen phase. At that point the quenched average \eqref{eq:QuenchedAv} becomes equivalent to the annealed one, meaning that we can replace $\langle \log \Gamma_R \rangle$ by $\log \langle \Gamma_R \rangle$. This yields the decay rate:
\begin{equation}
\gamma= -\lim_{R \to \infty} \frac{\log \langle \Gamma_R \rangle}{R}=-\log \quadre{k \int d\epsilon \, f(\epsilon) \Big| \frac{J_\perp}{\epsilon}\Big|^2} 
\end{equation}
which equals $\psi(\eta=1)$.

 \section{Details of the evaluation of $\gamma_L$ }\label{app:Calculation}
In this appendix we provide details on the explicit evaluation of the decay constant $\gamma_L$ in a liquid environment, the result of which is shown in Fig. \ref{fig:FigPathsMany} for a particular choice of parameters {($J_z=1= J_\perp$ and infinite temperature $\beta=0$)}. The main subtlety in the calculation consists in regularizing the integral:
 \begin{equation}\label{eq:OmegaNum}
\omega_\xi(\mathcal{E})= \int_{-\infty}^{\infty} dh  \, \frac{\text{exp} \tonde{-\frac{[h- \mathcal{E}]^2}{2 V_\beta}}}{\sqrt{2 \pi V_\beta}} e^{2 \xi \log  \left| \frac{J_\perp}{h} \right| },
 \end{equation}
which has a divergence for $\xi >1/2$. Following the original argument in \cite{anderson1958absence}, this is regularized by putting a cutoff around $h=0$ on a scale $\Delta$ representing the typical value of self-energy corrections. In our setting, the relevant scale is $\Delta \sim J_\perp^2/ \sqrt{J_z^2+ W^2} $. We therefore restrict the average in \eqref{eq:OmegaNum} to the domain $\mathbb{R}_\Delta= \mathbb{R}/[-\Delta, \Delta]$, and for simplicity, we neglect the correction to the normalization factor due to the cutoff $\Delta$. We now provide the explicit expression of the function $\Psi_L$ obtained with this regularization.

The effective locator becomes:
 {\medmuskip=0mu
\thinmuskip=0mu
\thickmuskip=0mu
  \begin{equation}
 \begin{split}
&\omega_\xi(\mathcal{E})= \frac{J_\perp^{2\xi}}{\sqrt{2 \pi  V_\beta}}  \int_{\mathbb{R}_\Delta} dh \,  \frac{ \text{exp} \tonde{-\frac{[h- \mathcal{E}]^2}{2 V_\beta}}}{h^{2 \xi} } =\frac{1}{\sqrt{\pi}}\tonde{\frac{J_\perp^2}{2 V_\beta}}^\xi \times \\
&\times \quadre{I_\xi\tonde{\frac{\mathcal{E}}{\sqrt{2 V_\beta}}, \frac{\Delta}{\sqrt{2 V_\beta}}} + I_\xi\tonde{-\frac{\mathcal{E}}{\sqrt{2 V_\beta}}, \frac{\Delta}{\sqrt{2 V_\beta}}} } \end{split}
 \end{equation}}
 where we introduced
  {\medmuskip=0mu
\thinmuskip=0mu
\thickmuskip=0mu
 \begin{equation*}
 I_\xi(a, u)= \int_u^\infty dh \,  \frac{  e^{-\tonde{h+a}^2}}{h^{2 \xi}}.
 \end{equation*}}
 A simple power expansion of the term $e^{- 2 h a}$ leads to the following representation:
  \begin{equation}\label{eq:ExpOmega}
 \begin{split}
&\omega_\xi(\mathcal{E})= \frac{2  f_\xi \tonde{\Delta} \,e^{-\frac{\mathcal{E}^2}{2 V_\beta}}}{\sqrt{\pi}}\tonde{\frac{J_\perp^2}{2 V_\beta}}^\xi \times\\
&\times \quadre{
 1+ \sum_{k=1}^\infty \frac{\Gamma \tonde{\frac{2k- 2 \xi+1}{2}, \frac{\Delta^2}{2 V_\beta}} }{f_\xi \, 2^{1-k} V_\beta^k   (2k)!}\mathcal{E}^{2k}},
 \end{split}
 \end{equation}
 where the coefficients of the expansion contain the incomplete Gamma function:
 \begin{equation}
 \Gamma(s,x)=\int_x^\infty t^{s-1}\, e^{-t} dt.
 \end{equation}
The term
 \begin{equation}
  f_\xi \tonde{\Delta}= \frac{1}{2}\Gamma \tonde{\frac{1}{2}-\xi, \frac{\Delta^2}{2 V_\beta}}
 \end{equation}
 is the source of the divergence as $\xi \to 1/2$ if no regularizer is present, $\Delta =0$. For $k \geq 1$ and $\xi \leq 1$ the coefficients of the expansion \eqref{eq:ExpOmega} remain regular for $\Delta \to 0$, and we therefore approximate them with their $\Delta \to 0$ limit. This allows us to re-sum the series into a function
  \begin{equation}
 \begin{split}
 &\Sigma_1(\mathcal{E}, \xi) \equiv \frac{1}{2 f_\xi(\Delta)}\sum_{k=1}^\infty  \frac{2^{k}\Gamma \tonde{k-\xi +\frac{1}{2} } }{ V_\beta^k  (2k)!}\,\mathcal{E}^{2k}=\\
 &\frac{2 \Gamma \tonde{\frac{3}{2}-\xi}\tonde{1- \,_1F_1\left(\frac{1}{2}-\xi ;\frac{1}{2};\frac{\mathcal{E}^2}{2
   V_\beta}\right)}}{ 2 f_\xi(\Delta) (2 \xi -1)},
   \end{split}
 \end{equation}
which remains regular when $\xi \to 1/2, \Delta \to 0$. 
Therefore:
    \begin{equation}\label{eq:OmegaMu}
 \begin{split}
[\omega_\xi]^\mu=[\lambda_\Delta(\xi)]^\mu  e^{- \frac{\mu \, \mathcal{E}^2}{2 V_\beta}} \quadre{
1+\Sigma_1(\mathcal{E}, \xi) }^\mu,
 \end{split}
 \end{equation}
 where
 \begin{equation}
 \lambda_\Delta(\xi)= \frac{2  f_\xi \tonde{\Delta} }{\sqrt{\pi}}\tonde{\frac{J_\perp^2}{2 V_\beta}}^\xi
 \end{equation}
accounts for the regularized singularity. With the aid of these formulas one obtains:
 \begin{equation}
 \begin{split}
 \Psi_L(\mu, \xi)&=-\frac{1}{\mu} \log \grafe{k \int d \mathcal{E} \rho(\mathcal{E})e^{- \frac{\mu \, \mathcal{E}^2}{2 V_\beta}} \quadre{
1+\Sigma_1(\mathcal{E}, \xi) }^\mu }\\
&- \log[\lambda_\Delta(\xi)],
 \end{split}
 \end{equation}
 which can be evaluated numerically as a function of $\mu, \xi$.
 
As discussed in the main text, determining the relevant value of the parameter $\xi$ requires to compute the sign of $x_{\rm SP}$ given in \eqref{eq:SaddleLambdaBmp}. Once \eqref{eq:OptimalDist} is used, Eq.~\eqref{eq:SaddleLambdaBmp} is equivalent to: 
 \begin{equation}\label{eq:Explicitx}
 x_{\rm SP}= - \frac{  \int d\mathcal{E}\,\rho(\mathcal{E}) [\omega_1(\mathcal{E})]^{\mu^*-1} \frac{\partial}{\partial z}\omega_z(\mathcal{E})\Big|_{z=1}}{ \int d\mathcal{E}\,\rho(\mathcal{E}) [\omega_1(\mathcal{E})]^{\mu^*}},
 \end{equation}
where explicitly 
 \begin{equation}
 \begin{split}
 &\frac{\partial \omega_z}{\partial z}\Big|_{z=1} = \log \tonde{\frac{J_\perp^2}{2 V_\beta}} \omega_1 +\frac{J_\perp^2}{ \sqrt{\pi} V_\beta}  \frac{\partial f_z(\Delta)}{\partial z}\Big|_{z=1} \omega_1\\
 &+\frac{J_\perp^2 f_1(\Delta)}{\sqrt{\pi} V_\beta} e^{-\frac{\mathcal{E}^2}{2 V_\beta}} \sum_{k=1}^\infty \tilde{\alpha}_k \mathcal{E}^{2k}
 \end{split}
 \end{equation}
with coefficients:
 \begin{equation}\label{eq:Der}
 \tilde{\alpha}_k= \frac{2^{k-1} }{ f_1(\Delta)\,V_\beta^k  (2k)!} \frac{\partial}{\partial z}\Gamma \tonde{\frac{2k-2 z+1}{2}, \frac{\Delta^2}{2 V_\beta}}\Big|_{z=1}.
 \end{equation}
It can be checked that upon taking the limit $\Delta \to 0$  of the coefficients of the series:
 \begin{equation}
 \begin{split}
 \Sigma_2(\mathcal{E}) \equiv \sum_{k=1}^\infty \tilde{\alpha}_k \mathcal{E}^{2k},
     \end{split}
 \end{equation}
the latter can be re-summed explicitly, too. Plugging the resulting expressions into \eqref{eq:Explicitx} and integrating over $\mathcal{E}$ allows one to evaluate explicitly $x_{\rm SP}$ as well. \\ 
The evaluation of the decay constant in a frozen environment, $\gamma_F$, is straightforward using the formula \eqref{eq:PolimeroFrozen}: to be consistent with the treatment in the liquid environment, we regularize the integral in \eqref{eq:PolimeroFrozen} by restricting the integration to the same domain $\mathbb{R}_\Delta$.

 \section{On the location of the path-freezing transition in a liquid environment}\label{app:NoMu1}
    In this appendix we show that the succession of phases and transitions as seen in Fig.~\ref{fig:FigPathsMany} for a system with a liquid environment is generic. In particular, this means that the path-unfreezing transition 
 (where $\mu^* \to 1$) always takes place within the intermittent metal phase, between the localization transition and the transition toward the non-resonating insulator. This follows indeed from simple convexity arguments.  

We start by showing that the path-freezing transition always occurs within the intermittent metal phase. This is equivalent to stating that the non-resonating insulator is always path-frozen. 
We show this by way of contradiction: Deep in the insulator phase, we are in the non-resonating phase, dominated by a single path and $\mu^*<1$. We now assume that for some parameters the solution of \eqref{eq:PolymerDerivative22} reaches $\mu^*=1$, while we are still in the non-resonating insulator, as defined by the condition that the dominating fluctuations have a positive decay rate, $x_{\rm SP}>0$, which requires to set $\xi=1$ in $\Psi_L(\mu,\xi)$. Under these assumptions, using \eqref{eq:SaddleLambdaBmp} together with \eqref{eq:OptimalDist} one obtains the expression: 
 \begin{equation}\label{xspa}
\begin{split}
 x_{\rm SP}= -\frac{\int d\mathcal{E}\, \rho(\mathcal{E}) \quadre{\int dh\, P(h| \mathcal{E}) \Big| \frac{J_\perp}{h}\Big|^2 \, \log \Big| \frac{J_\perp}{h}\Big|^2 }}{
 \int \, d \mathcal{E} \rho(\mathcal{E}) \,  \omega_1(\mathcal{E})},
 \end{split}
\end{equation}
with $\omega_1(\mathcal{E})$ defined in Eq.~(\ref{eq:OmegaXi}).
 For a generic convex function $\mathcal{F}$, the following inequality holds:
 \begin{equation}\label{eq:Convex}
 \int dh\, P(h|\mathcal{E})\, \mathcal{F}(g(h)) \geq \mathcal{F} \tonde{ \int dh\, P(h|\mathcal{E})\,g(h)},
 \end{equation}
 which is a strict inequality for $\mathcal{F}$ strictly convex (provided $g(h)$ is not a constant or, equivalently, that the density $ P(h|\mathcal{E})$ does not collapse to a delta distribution).
Exploiting this for $\mathcal{F}(y)= y \log(y)$ {(which is strictly convex for $y>0$)} and $g(h)= |J_\perp/h|^2$, we obtain
\begin{equation}
\label{xSPappC}
\begin{split}
&- x_{\rm SP}\\
& >  \frac{  \int d\mathcal{E}\, \rho(\mathcal{E}) \quadre{\int dh\, P(h| \mathcal{E}) \Big| \frac{J_\perp}{h}\Big|^2}  \log \quadre{ \int dh\, P(h| \mathcal{E}) \Big| \frac{J_\perp}{h}\Big|^2 }}{\int \, d \mathcal{E} \rho(\mathcal{E}) \,  \omega_1(\mathcal{E})}\\
&= \frac{  \int d\mathcal{E}\, \rho(\mathcal{E}) \omega_1(\mathcal{E})  \log \quadre{\omega_1(\mathcal{E})  }}{\int \, d \mathcal{E} \rho(\mathcal{E}) \,  \omega_1(\mathcal{E})},
\end{split}
\end{equation}
The derivative in \eqref{eq:PolymerDerivative22} can be written as:
\begin{equation}\label{eq:stat3} 
\frac{\partial \Psi_L (\mu, \xi)}{\partial \mu}= \frac{1}{\mu} \tonde{\Psi_L (\mu, \xi)-\frac{\int d\mathcal{E} \rho (\mathcal{E}) \log[ \omega_{\xi}] \omega_{\xi}^{\mu}}{\int d\mathcal{E} \rho (\mathcal{E}) \omega_{\xi}^{\mu}}}.
\end{equation}
Under our assumptions, $\mu^*=1$ is a solution of $\partial \Psi_L/ \partial \mu=0$ with $\xi=1$, thus:
\begin{equation}\label{eq:PolymerDerivative}
\Psi_L (1, \xi)- \frac{\int d\mathcal{E} \rho(\mathcal{E}) \, \omega_{\xi} \log \omega_{\xi} }{\int \, d \mathcal{E} \rho(\mathcal{E}) \,  \omega_{\xi}} \Big|_{\xi=1}=0.
\end{equation}
Using that the liquid decay rate satisfies $\gamma_L= \Psi(\mu^*, \xi^*)$ (with here $\mu^*=\xi^*=1$), together with Eq.~(\ref{xSPappC}) we obtain
\begin{equation}\label{eq:Ineq1}
\begin{split}
&- x_{\rm SP}> \gamma_L > 0,
\end{split}
\end{equation}
since everywhere in the insulating phase $\gamma_L>0$. This is
a contradiction to our starting assumption that $x_{\rm SP}$ was non-negative.  {Hence, we conclude that being in the non-resonating insulator ($x_{\rm SP} > 0$) is incompatible with reaching the path unfreezing transition ($\mu^*\to 1$), which can thus only occur in the intermittent metal phase.} A simple extension of these arguments to an arbitrary $\mu^* \neq 1$ satisfying \eqref{eq:PolymerDerivative22} shows that the delocalization transition ($\gamma_L \to 0$) can only occur within the intermittent metal phase ($\xi<1$). Namely,  imposing $\gamma_L =0$, while fixing the parameter $\xi$ to $\xi=1$, implies that the decay constant $x_{\rm SP}$ as defined in Eq.~\eqref{xspa} becomes negative and thus is unphysical. \\

Let us now turn to the localization transition, and show that path-unfreezing (as signaled by $\mu^*\to 1$) always precedes it within the insulator. Technically, we need to show that, upon approaching delocalization the maximizing $\mu^*$ reaches 1 before $\gamma_L \to 0$. This will imply that exactly at the transition, $\mu^*$ can no longer be chosen as a maximizer of $ \Psi_L (\mu, \xi)$ with $\mu^*<1$; instead, the maximum of $\Psi_L$ has to be assumed on the boundary, $\mu^*=1$. 

Let us now show that $\mu^*\to 1$ strictly within the insulator. Within the intermittent metal phase  $\xi^*$ is determined by the condition (cf. Eq.~(\ref{eq:KappaStar}):
\begin{equation}\label{eq:XZ}
\int d\mathcal{E} \rho (\mathcal{E}) \omega_{\xi^*}^{\mu^*-1} (\mathcal{E})\quadre{\int dh P(h|\mathcal{E}) \Big| \frac{J_\perp}{h}\Big|^{2 \xi^*} \log \Big| \frac{J_\perp}{h}\Big|^2 }=0.
\end{equation}
The inequality \eqref{eq:Convex} applied to $g(h)=|J_\perp/h|^{2 \xi^*}$ and $\mathcal{F}(y)= y \log y$ (for $y>0$) gives:
\begin{equation}
\int dh P(h|\mathcal{E}) \Big| \frac{J_\perp}{h}\Big|^{2 \xi^*} \log \Big| \frac{J_\perp}{h}\Big|^{2\xi^*} > \omega_{\xi^*}(\mathcal{E})  \log [\omega_{\xi^*}(\mathcal{E})],
\end{equation}
which upon injection into Eq.\eqref{eq:XZ} implies:
\begin{equation}\label{eq:XZY}
\int d\mathcal{E} \rho (\mathcal{E}) \omega_{\xi^*}^{\mu^*} (\mathcal{E})  \log [\omega_{\xi^*}(\mathcal{E})]<0.
\end{equation}
From the expression \eqref{eq:stat3} for the derivative $\partial \Psi_L/\partial\mu$ we conclude that the value of $\mu^*$ for which  $\gamma_L= \Psi_L (\mu^*, \xi^*)=0$ can not be a stationary point of $ \Psi_L$, since \eqref{eq:XZY} implies:
\begin{equation}
\frac{\partial \Psi_L (\mu, \xi)}{\partial \mu} \Big|_{\mu^*, \xi^*}=-\frac{1}{\mu^*} \frac{\int d\mathcal{E} \rho (\mathcal{E}) \log[ \omega_{\xi^*}] \omega_{\xi^*}^{\mu^*}}{\int d\mathcal{E} \rho (\mathcal{E}) \omega_{\xi^*}^{\mu^*}} >0.
\end{equation}
Therefore, at the delocalization transition $\mu^*$ must lie at the right boundary of the admissible interval $0\leq \mu\leq 1$, i.e.,  $\mu^*=1$. From this it follows that the delocalized phase is always adjacent to a path-unfrozen, intermittently metallic insulating phase.

\bibliographystyle{unsrt}
\bibliography{bibRefTdep}

\end{document}